\documentclass[useAMS,usenatbib]{mn2e}
\usepackage[pdftex]{graphicx}
\DeclareGraphicsExtensions{.png,.pdf,.jpg}
\usepackage{float}
\restylefloat{figure}
\usepackage{caption}
\usepackage{subcaption}
\usepackage[authoryear]{natbib}
\usepackage{color}
\usepackage{amsmath}

\title[3D HD models of H1-67]{3D hydrodynamical models of point-symmetric planetary nebulae:  \bf{the special case of H\,1-67}\thanks{Based upon observations carried out at the Observatorio Astron\'omico Nacional on the Sierra San Pedro M\'artir (OAN-SPM), Baja California, M\'exico.}}
\author[Rechy-Garc\'ia, J. S., et al]{Rechy-Garc\'ia, J. S.$^{1}$\thanks{E-mail: jrechy@astro.unam.mx}, Pe\~na, M.$^{1}$, Vel\'azquez, P. F.$^{2}$, \\
\\
$^{1}$Instituto de Astronom\'ia, Universidad Nacional Aut\'onoma de M\'exico, Apdo. Postal 70264, 04510, Ciudad de M\'exico, M\'exico. \\
$^{2}$Instituto de Ciencias Nucleares,  Universidad Nacional Aut\'onoma de M\'exico, Apdo. Postal 70543, 04510, Ciudad de M\'exico, M\'exico. }
\begin{document}

\date{}

\pagerange{\pageref{firstpage}--\pageref{lastpage}} \pubyear{2002}

\maketitle

\label{firstpage}

\begin{abstract}
{We present 3D hydrodynamical simulations of a precessing jet with a time-dependent ejection velocity or a time-dependent ejection density, interacting with a circumstellar medium given by a dense, anisotropic, and slow AGB wind, forming a torus. We explore a set of configurations with different values for the precession angle and number of ejections. The temporal evolution of these models is analised at times up to 1500  or 1800 yr.   
From our hydrodynamical models, we obtain position-velocity diagrams (PV diagrams) in the [N\,{\sc ii}] $\lambda$6583 line to be compared with high resolution observations of the planetary nebula H\,1-67. From spectral data this object shows high-velocity jets and a
point-symmetric morphology. With our synthetic PV diagrams we show that a precessing jet with a time-dependent ejection velocity or a time-dependent ejection density reproduce the point-symmetric morphological structure for this nebula if the precession cone angle is larger than 30$^{\circ}$.
Our synthetic PV diagrams can be used to understand how the S-like morphology, also presented  by other planetary nebulae, is formed. For H\,1-67 we found a heliocentric velocity of -8.05 km  s$^{-1}$ and height below the galactic plane of $-$451.6 pc.
}
\end{abstract}

\begin{keywords}
Planetary nebulae: general --- hydrodynamics --- methods: observational ---
methods: numerical --- ISM: jets and outflows.
\end{keywords}

\section{Introduction}
Planetary nebulae (PNe) are expanding regions of ionized gas, whose central stars had initial masses lying in the range of 0.8 - 8 M$_\odot$. 
In a traditional scenario, PNe are formed in a phase of mass loss 
of stars that are leaving the Asymptotic Giant Branch (AGB). 
These stars evolve towards high effective temperatures at about constant luminosity and ionize the ejected shell, afterwards they become a white dwarf (WD).

It is well known that PNe have a wide variety of shapes, 
showing shells, jets, torii, knots or internal structures and they are classified
according to their morphology \citep[]{Balick1987, Parker2006, SahaiMorris2011}. The main morphological groups are: round, elliptical, bipolar, and multipolar.
Some PNe show a jet-like morphology which can be indicative of a PN evolving from a binary system \citep[]{SokerLivio1994, Bond2000, DeMarco2009, SokerRappaport2000}.

In this morphological scheme PNe catalogued as point-symmetric were  introduced by \citet{SchwarzCorradi1993} and \citet{StanghelliniCorradi1993}. These are defined 
as PNe whose morphological components show point reflection symmetry with respect to the centre,
similar to a S-shape. Examples of point-symmetric PNe  are IC\,4634
\citep{GuerreroMiranda2008}, NGC\,6309 \citep{VazquezMiranda2008}, and others. A very interesting case is the PN Fleming 1 (Fg\,1) which shows a very extended S-shape where the filaments are moving at a  high velocity of $\pm$ 75 km s$^{-1}$ \citep{LopezMeaburn1993}.
See the catalogue by \citet{SahaiMorris2011} and the compilation by \citet{Guerrero2000}  for a list of point-symmetric PNe.\\
Point-symmetric PNe have been studied in previous works \citep[]{MirandaSolf1992, LopezMeaburn1993}.
They are situated at scale height over the galactic plane of 310 pc \citep{GarciaSeguraFranco2002} which is smaller than the general scale height of disk PNe.
It has been suggested that the formation of this morphology is a consequence of a precessing collimated outflow \citep{Guerrero2000} and a binary system \citep{LivioPringle1996}.\\
Some models have been produced to understand this kind of morphology \citep[]{GarciaSeguraLopez2000, RijkhorstIcke2004, RagaCanto1993, RagaEsquivel2009}. 
\citet{Cliffe1995} calculated a hydrodynamical model for the formation of point-symmetric PNe and they concluded that the knots seen in these PNe can be interpreted as
resulting from the periodic ejection of dense material by a precessing jet.
3D numerical simulations like
the ones by \citet{HaroCorzoVelazquez2009} have shown that a jet ejected from a  binary system can lead to morphological structures of point-symmetric PNe.

The ideas described above have been succesfully applied to, e.g., the well known point symmetric PN IC\,4634 by \citet{GuerreroMiranda2008}, who from very deep HST images and high-resolution spectroscopy, and by using the code Yguaz\'u (described in \S 2.1), constructed a hydrodynamical model, which included  a precessing fast collimated outflow interacting with nebular material. Due to they found low radial velocities of the S-shape structure ($\pm$ 20 km s$^{-1}$), these authors have assumed that  the precession axis is on the plane of the sky. 

Other well known point symmetric nebulae such as NGC\,6309 and Fg\,1 have been studied with deep images and kinematical models. It is worthy to say that the extended S-shaped Fg\,1  resulted to possess a binary  stellar system which is the most probable cause for its morphology \citep{BoffinMiszalski2012}. A  close binary nucleus has also been found in the spectacular quadrupolar PN NGC\,5189. This binary system is composed by a hot [WO\,1] star and a massive WD  \citep{ManickMiszalski2015}. The nebula contains two central toroidal structures (one IR and the other optical) associated to bipolar outflows and it shows multiple point-symmetric low-ionization knots or ansae and collimated outflows. \citet{SabinVelazquez2012} found moderate expansion velocities  in the three bubbles conforming the nebula, but the largest velocities found at both extremes of the nebula are about $-$40 km s$^{-1}$ and 50 km s$^{-1}$. Other PNe with precessing collimated outflows are Hen\,2-1, Hen\,2-141, Hen\,2-429, Hu\,2-1, J320, M\,1-61, NGC\,5307, NGC\,6881, PC19 among others.
For none of these objects reported as point-symmetric PN,  except IC\,4634, a specific hydrodynamical  model has been constructed. 
\\

\begin{figure}
\begin{center}
\includegraphics[width=0.8\linewidth]{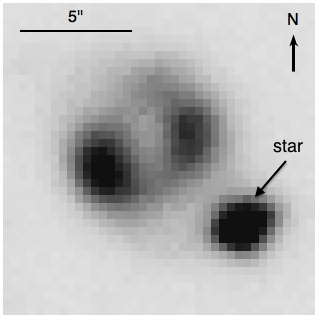}
  \caption{ H$\alpha$ image of PN H\,1-67, obtained that OAN-SPM using the MES spectrograph. North
  is up and East to the left. The exposure time is 100 s. We observe a face-on torus, with two bright condensation on the east and west side.}
  \label{fig:H167image}
 \end{center}
\end{figure}

\begin{figure}

\includegraphics[width=1.05\linewidth]{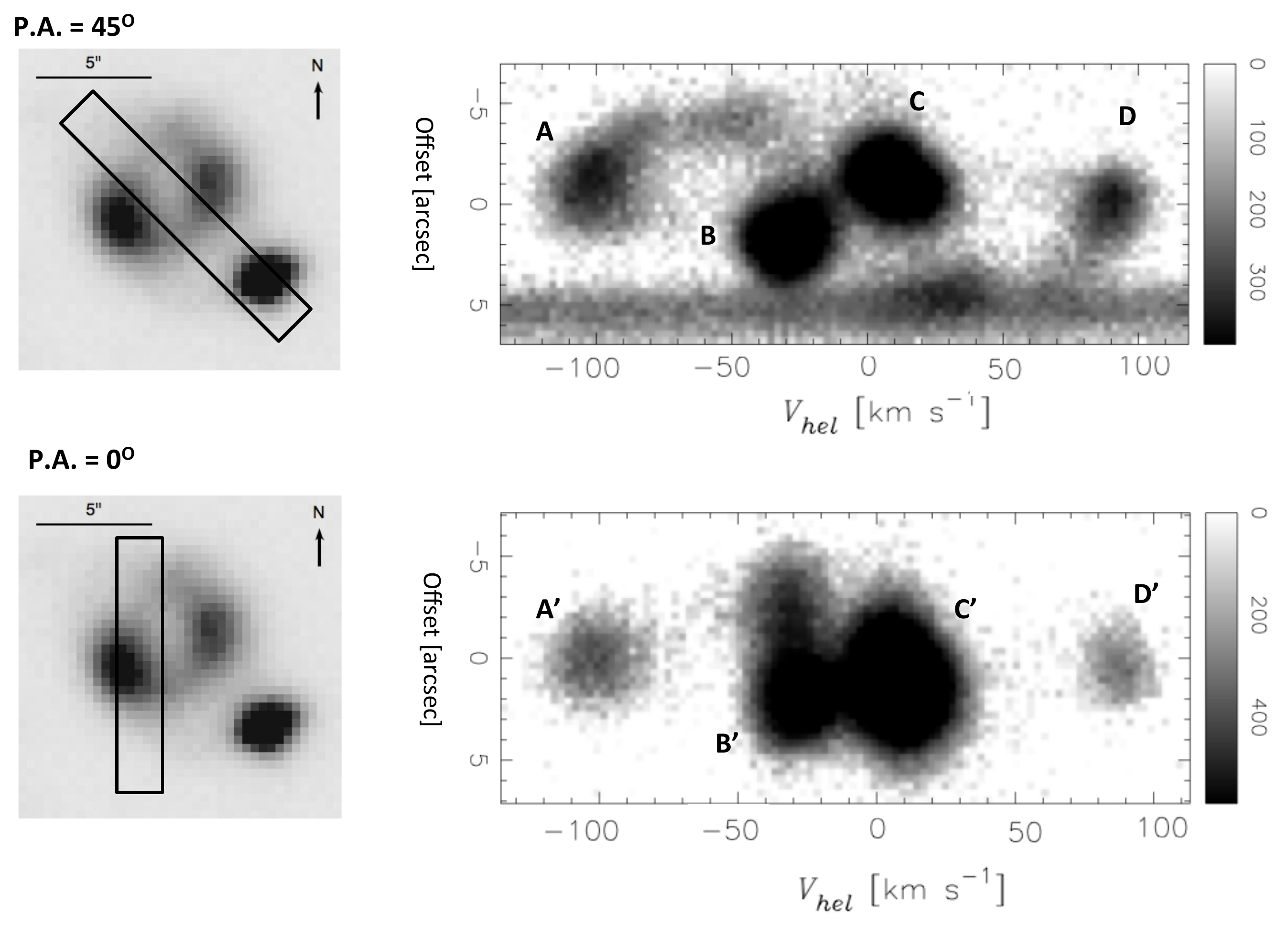}
  \caption{ Position-velocity diagrams of H\,1-67 in the [N\,{\sc
        ii}]$\lambda$6583 line,  for two positions of the
    slit (shown in the left frames). The slit to the East (bottom figure) is located 
    at about 1.7 arcsec  from
    the centre of the object. The bright condensations (marked as B, C and B', C' ) in the right panel  correspond to the toroidal component. The wide faint components come from a bipolar
    precessing jet ending in knots (marked as A, D and A', D'). The grey scale to the right is the intensity in counts. The ratio between the brightest and faintest regions is of the order of 20. }
  \label{fig:H1-67pv}
\end{figure}

In this and in our previous work, our aim has been to analyse compact PNe presenting  round or toroidal images but showing high velocity ejections larger than 70 km s$^{-1}$ observed in Position-Velocity diagrams. Due to their  appearance, these objects  have been classified as `compact' or `elliptical' PNe. The objects in our analysis have been selected from the San Pedro M\'artir (SPM) Kinematical Catalogue of Galactic  Planetary Nebulae \citep{Lopez2012}.\\
By analysing position velocity (PV) diagrams of high resolution spectra of PNe that in images appear compact with no evident structure, we have found that some of them show high velocity ejections that seem to correspond to collimated material thrown  through the poles of a torus.   Such are the cases of M\,1-32 and M\,3-15 \citep{RechyGarcia2017}. Other objects show point-symmetric features at high velocities. \\
One of the most interesting cases is the PN H1-67, presented in Figs. \ref{fig:H167image} and  \ref{fig:H1-67pv}. The image and spectra were obtained with the  Manchester Echelle  Spectrograph (MES) attached to the 2.1-m telescope at Observatorio Astron\'omico Nacional, San Pedro M\'artir (OAN-SPM), B.C., M\'exico, on the night 2013 May 18. A detailed analysis of these data is presented in \S 4. Here we want to show that this PN looks compact in direct images (see Fig. \ref{fig:H167image} and also the image presented by Reese \& Zijlstra  2013), showing a relatively face-on ``broken" torus with two bright condensations on the East and West sides, while the PV diagram (Fig. \ref{fig:H1-67pv} top) obtained from the spectrum with the slit crossing the nebula through the central position, along the symmetry  axis, at P.A. = 45$^o$, shows two central bright condensations co\-rres\-pon\-ding to the broken torus and a S-shape ejection ending in two knots at about $\pm$100 km s$^{-1}$. This velocity structure corresponds to a point-symmetric nebula, fact that is not noticed in the direct images. In addition, the PV diagram obtained with a slit crossing at P.A. = 0$^o$ shows the condensations corresponding to the broken torus and two knots at velocity of about $\pm$100 km s$^{-1}$ (Fig. \ref{fig:H1-67pv} bottom). No S-shape is noticeable in this latter spectrum.

From the SPM Kinematical catalogue mentioned above, other compact nebulae such as M\,2-36 and Wray 16-411, presented in \S 5, show similar S-shape velocity structures.

The main aim in the present study is to  examine the formation of such point-symmetric morphologies. This can be carried out by analising the history of the injected gas, which  shapes this kind of nebulae.  
Following \cite{Cliffe1995} we will study the effect of a periodic ejection of dense material in the form of a precessing jet,  
 by means of hydrodynamical models. We calculate numerical simulations of a jet  that changes its direction describing a precession cone in order to
reproduce the S-like morphology and we include a time-dependent ejection velocity or a time-dependent ejection density, in order to explain the observed structures in these PNe. Furthermore, the jet interacts  with a ``clumpy" slow wind, equatorially concentrated emulating the presence of a broken dense torus.

This paper is organized as follows, we describe the hydrodynamical models and the initial settings in \S 2. In \S 3 the different morphologies obtained from the models 
are presented. In \S 4 we compare our results with observations of PN H\,1-67 that presents S-like morphological features and compute specific models for this object. Other similar objects are presented in \S 5 and our conclusions are summarised in \S 6.

\section{Numerical simulations}
\subsection{The hydrodynamical model and the Yguaz\'u code}

In order to reproduce the observational S-shape PV diagrams, we have carried out
3D hydrodynamical simulations taking into account the observed features of H1-67 described above:  a broken dense
torus and several clumps, at high and medium velocities, distributed along a filament with a S-like morphology. 

 From the H$\alpha$ image it is evident that H\,1-67 have a dense `broken''
torus and that the high velocity material flows through the poles of the torus. 
In order to calculate a hydrodynamical model for this object, we have considered a circumstellar medium
 with a density distribution given by a dense and low-velocity AGB wind forming a torus. Subsequently, a second
outflow is launched by the central star. Different types of second outflow were tested: an isotropic wind,
an anisotropic wind (with anisotropical distributions in velocity or density), a cylindrical jet and a conical jet. 
The model that best explains the PV diagrams observed for
 H\,1-67 consists of two components: a dense torus and a cilindrical jet. 

Additionally, and in order to simulate a broken dense torus, we have built a clumpy wind by modulating the density by a fractal structure with a spectral index of 11/3, which generates a noisy structure and it is consistent with a turbulent interstellar medium \citep{EsquivelLazarian2003}. We imposed this clumpy density on the initial condition
and this fluctuates with a 20 per cent of the mean density value.

Finally, to obtain the observed distribution of the clumps in the PV diagrams, a precession and a velocity ejection variability of the outflow were introduced.  
Alternatively, a model including a time-dependent ejection density  of the outflow  has been computed and it will be described  ahead.

The numerical simulations were performed
with the hydrodynamical code Yguaz\'u \citep{Raga2000}. This is a
3D adaptive code which solves the gasdynamic equations employing the
``flux vector splitting" scheme of \citet{VanLeer1982}. 
Together with these
equations,  a system of equations for atomic/ionic
species are also integrated in order to provide a cooling function. For these simulations, we have used
the total abundances of N, O and S (relative to hydrogen), obtained
by \citet{EscuderoCosta2004} for H1-67, which are similar to the ones presented by many PNe in the Galaxy.

\subsection{Initial setup}

Our numerical simulations use an adaptive
Cartesian grid. We computed several models varying the parameters of the outflow. 
The sizes of the computational domain are of $10^{18}$ cm along
the x- and y-directions, and 2$\times$10$^{18}$ cm along the z-axis for model 1, while for models 2, 3 and 4 (the latter one with a jet with density variations), the size along the y-axis increases to 2$\times$10$^{18}$ cm (see Table 1).\\ 
At the beginning, following the work by \citet{Mellema1991},  to form the torus we have imposed on all computational domain the density distribution of an anisotropic AGB wind,  which is given by:

\begin{equation}
\rho (r, \theta) = \rho_{0} g (\theta) (r_{0}/r)^{2},
\label{rhor}
\end{equation}

\noindent where $r$ is the distance to the centre of computational domain, $\theta$ is the polar angle measured with respect to the z-axis, and $\rho_0$ is the density for a reference radius $r_0$, which is given by:
\begin{equation}
\rho_0 = \mathrm{\dot M_{AGB} }/ (4 \pi r_{0}^2 V_{AGB}) , 
\end{equation}
being $\mathrm{\dot M_{AGB}}$  the mass loss rate  
and $V_{AGB}$ the terminal wind velocity, which was set to 20 km s$^{-1}$.  The angular dependence of the wind is described by the following equation:
 \begin{equation}
  g(\theta) = 1 - A \left[\frac{1-\exp(-2 B \cos^{2} \theta)}{1-\exp(-2 B)}\right],
  \label{g}
\end{equation}
where $A$ is the parameter which determines the equator to pole density ratio, and $B$ gives the variation of the density from
the equator ($xy$ plane or $\theta=90^{\circ}$) to the poles ($+z$ direction or $\theta=0^{\circ}$; $-z$ direction or $\theta=180^{\circ}$).
\begin{figure}
\centering
   \includegraphics[width=0.6\linewidth]{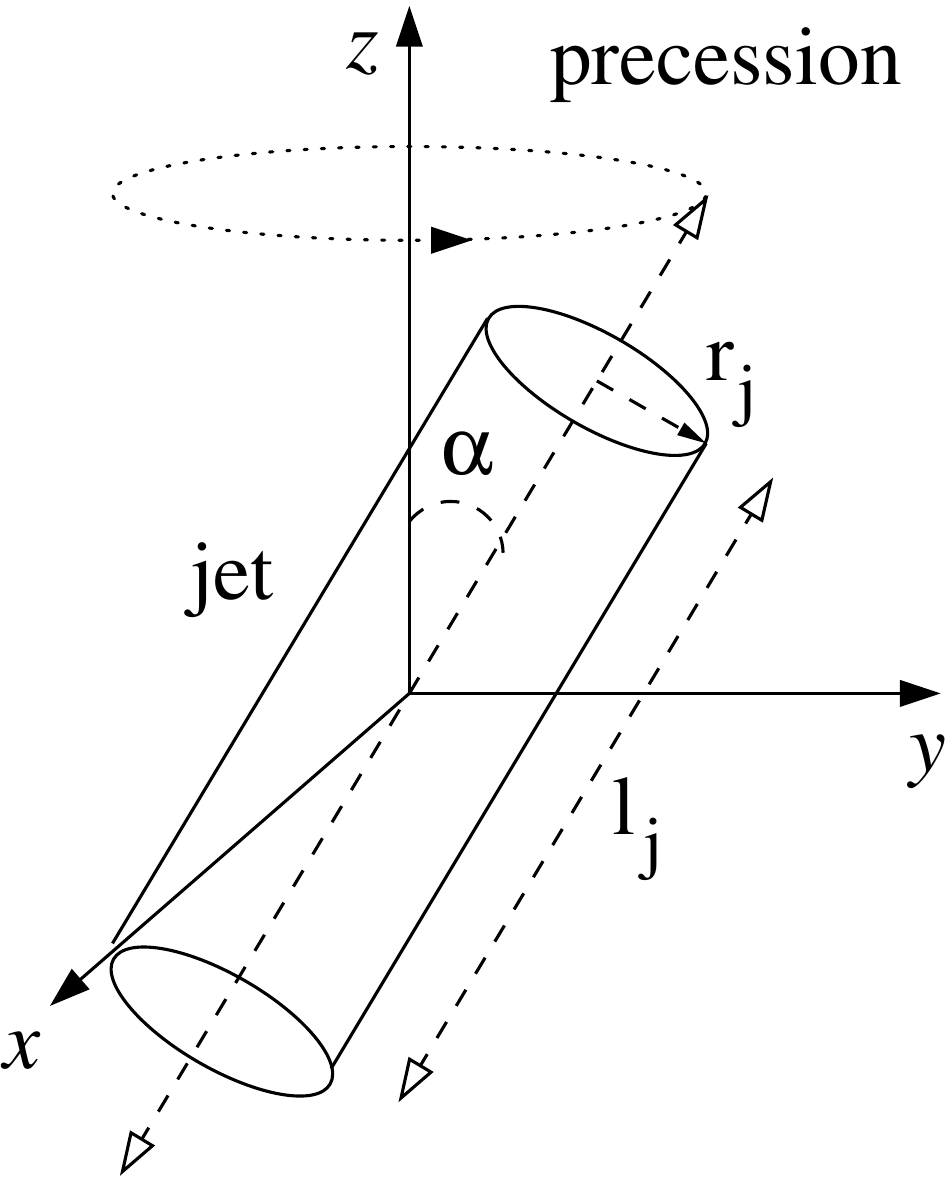}
   \caption{Scheme of the jet configuration}
   \label{fig:jet}
\end{figure}

A jet is imposed at the centre of the computational domain in a cylindrical volume with radius $r_j=r_0$ and a length $l_j=2 r_0$ (see Fig. \ref{fig:jet}). The jet direction describes a precession cone, with a semi-aperture angle $\alpha$ and a period $\tau_p = \tau_{ch}/f_p$, where $f_p$ is a constant representing the factor of precession, and 
$\tau_{ch}$ is a characteristic time given by:
\begin{equation}
\tau_{ch}=\frac{0.5~z_{max}}{v_0(1+\Delta v) \cos\alpha}
\label{tauchar}
\end{equation}
Furthermore, the jet velocity depends upon the time as:
\begin{equation}
v_j=v_0 \bigg(1+\Delta v \cos\big(2\pi \frac{t}{\tau_v}\big)\bigg)
\label{vjet}
\end{equation}
where $v_0$ is the mean jet velocity, $\Delta v$ is the amplitude of the velocity variation, $t$ is the time, and $\tau_v$ is the period of the velocity variation, which is given by $\tau_v=\tau_{p}/f_q$, being $f_q$ a constant (the values of $f_p$ and $f_q$ used in our simulations are listed in Table 1). This jet has a constant density of 150 cm$^{-3}$.

In the case of a jet with a time-dependent density ejection, the jet velocity is fixed to $v_0$ and the density is given by: 
\begin{equation}
\rho_j=\rho_{j0} \bigg(1+\Delta \rho_j \cos\big(2\pi \frac{t}{\tau_d}\big)\bigg)
\label{djet}
\end{equation}
where $\rho_{j0}$ is the mean density of the ejection, $\Delta\rho_j$ is the amplitude of the density variation, $t$ is the time and $\tau_d$ is the period of the density variation.  In this case, $\tau_{ch}$ in Eq. \ref{tauchar} changes accordingly.

In all simulations the values of parameters $A$ and $B$ in Eq.(\ref{g})  were chosen as 0.99 and 5.0, respectively, because these values are adequate to produce a highly- contrasted thin disk \citep{Mellema1991}. 
Such a disk (torus)  constraints the stellar ejection helping to produce a bipolar nebula. In this torus the flow to external  density ratio is much smaller than one in the equator ($\theta$ = 90$^\circ$ in Eq.(\ref{g})) therefore the injected gas rapidly slows down and the ejection of the jet is prevented in this direction. Instead, this ratio is about 2 when $\theta$ = 0$^\circ$ (pole direction), therefore the jet escapes easily. We have also explored a model with a thick torus (with B=2) finding that in this case the jet is slowed, being necessary to employ larger integration times. However, this
model is not successful in reproducing the ``S" morphology observed in the PV diagrams.

The temperature for
the AGB wind was fixed as 1000 K while its $\mathrm{\dot M_{AGB}}$ was of $5\times 10^{-6} \mathrm{M_{\odot}\ yr^{-1}}$. The value of $r_0$ was chosen as 5 pixels of the grid at its maximum resolution, i.e., $3.08\times 10^{16}$~cm. 
For the jet with velocity variability we have set $v_0$ and $\Delta v$ as 140 km s$^{-1}$ and 0.5, respectively, while for the jet with density variability $\rho_{j0}$ is 150 cm$^{-3}$ and $\Delta \rho_j$ is 0.7. The variability period is the same in both cases.

Initially we computed a model with a precessing cone with a semi amplitude angle $\alpha$ equal to 15$^{\circ}$. Such a model did not produce a S-shape in the PV diagrams of the ejection and we discarded it. This model  is not presented here, but helped us to understand that a  larger $\alpha$ is required. Afterwards we computed models with $\alpha$ of 30$^{\circ}$ and 40$^{\circ}$ which  show S-shape structures in the PV diagrams.

Different values of $\alpha$, $f_q$ and $f_p$ were used (see Table 1)  to analise the PV diagrams produced.  In the following we describe the properties of Model 1  with $\alpha$= 30$^{\circ}$ , $f_p$=4 (the jet precess 4 times), $f_q$=2 (two ejections per precession)   and Model 2 with $\alpha$=40$^{\circ}$, $f_p$=1.5  and $f_q$=2.

\begin{table*}
\centering
 \begin{minipage}{120mm}
  \caption{Input for hydrodynamical simulations}
\begin{tabular}{lllccccc}
\hline \hline
Model & Computational domain&$f_{p}$&$f_{q}$& $\alpha$ & $\tau_v$ &$\tau_{ch}$\\
& $x-$ , $y-$, and $z-$axis, (cm) &&&& yr& yr \\
\hline

model 1 & (1.05, 1.05, 2.10)$\times$ 10$^{18}$  &4&2&30$^{\circ}$ & 230 & 1830\\

model 2 & (1.05, 2.10, 2.10)$\times$ 10$^{18}$  &1.5&2&40$^{\circ}$ & 690 & 2070\\

model 3 & (1.05, 2.10, 2.10)$\times$ 10$^{18}$  &1.5&4&40$^{\circ}$ & 345 & 2070\\

model 4$^a$ & (1.05, 2.10, 2.10)$\times$ 10$^{18}$  & 1.5&4&40$^{\circ}$ & 304 & 1826 \\
\hline
\multicolumn{5}{l}{$^a$ model with variable density. In column 6 $\tau_d$ is presented.}
\end{tabular}
\label{tab:hydro-model}

\end{minipage}
\end{table*}

\section{Numerical results of hydrodynamical models}

\subsection{Density stratification evolution}

\begin{figure}
\centering

   \includegraphics[width=1\linewidth]{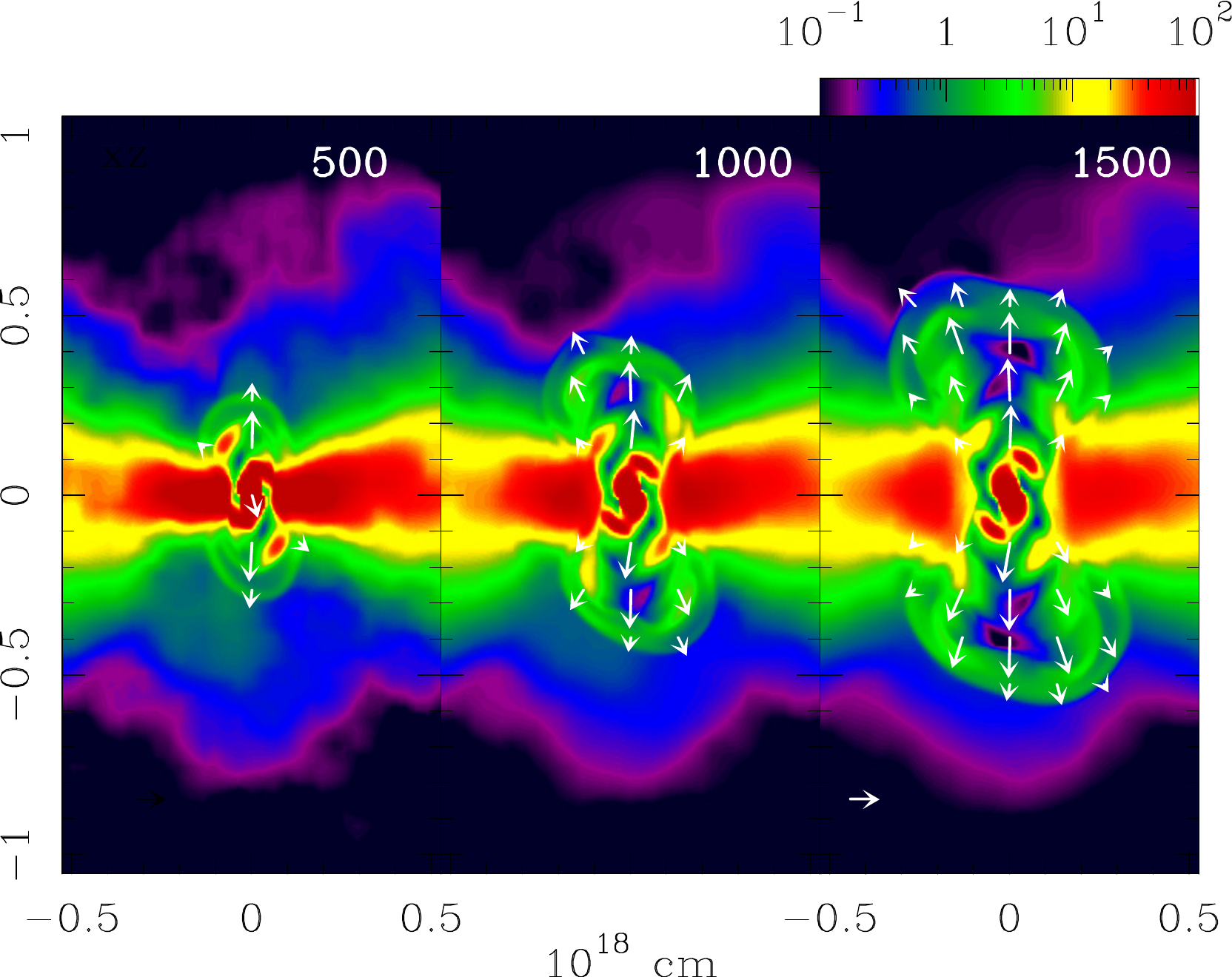}
   
   \includegraphics[width=1.0\linewidth]{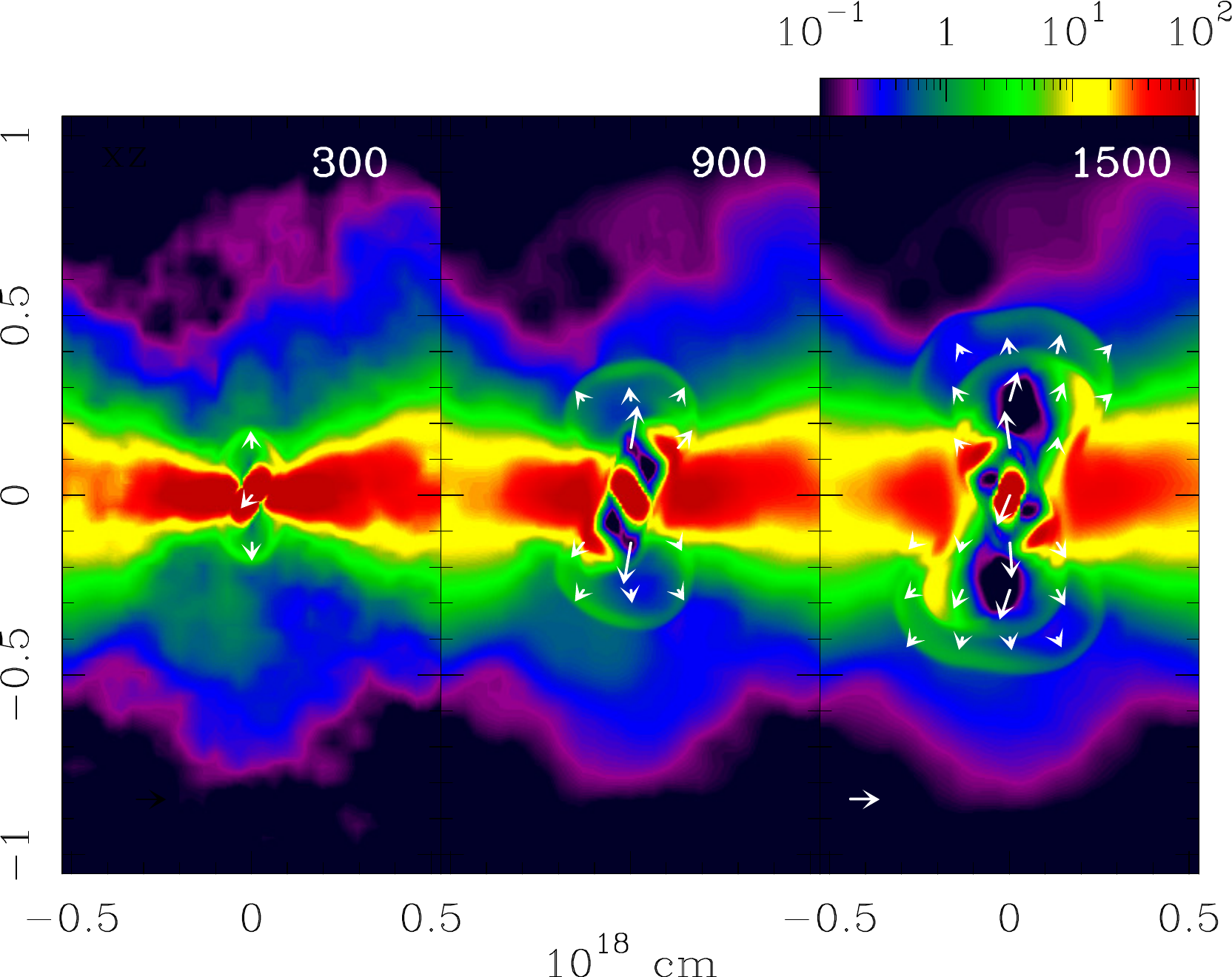}
 \caption{Electron density distribution maps ($xz-$projection),
    corresponding to models 1 ($\alpha$=30$^\circ$, upper panel) and  2 ($\alpha$=40$^\circ$, bottom panel), at different evolutionary times (in yr). 
Both axes are given in units of
   10$^{18}$ cm and the logarithmic color scale gives the number density in units of cm$^{-3}$.
     The arrows show the velocity field of the jet material. The small arrow at the bottom right frame corresponds to a velocity of 100 km s$^{-1}$.}
  \label{fig:velDenAlfa5cm}
\end{figure}

Fig. \ref{fig:velDenAlfa5cm} displays the temporal evolution of the electron density stratification at integration times of 500, 1000, and 1500  yr for models 1 and 2 mentioned above. These maps show density distribution projected on the xz-plane. The white arrows indicate the velocity field of the material in the jet.  

\begin{figure}
\centering

 \includegraphics[width=1\linewidth]{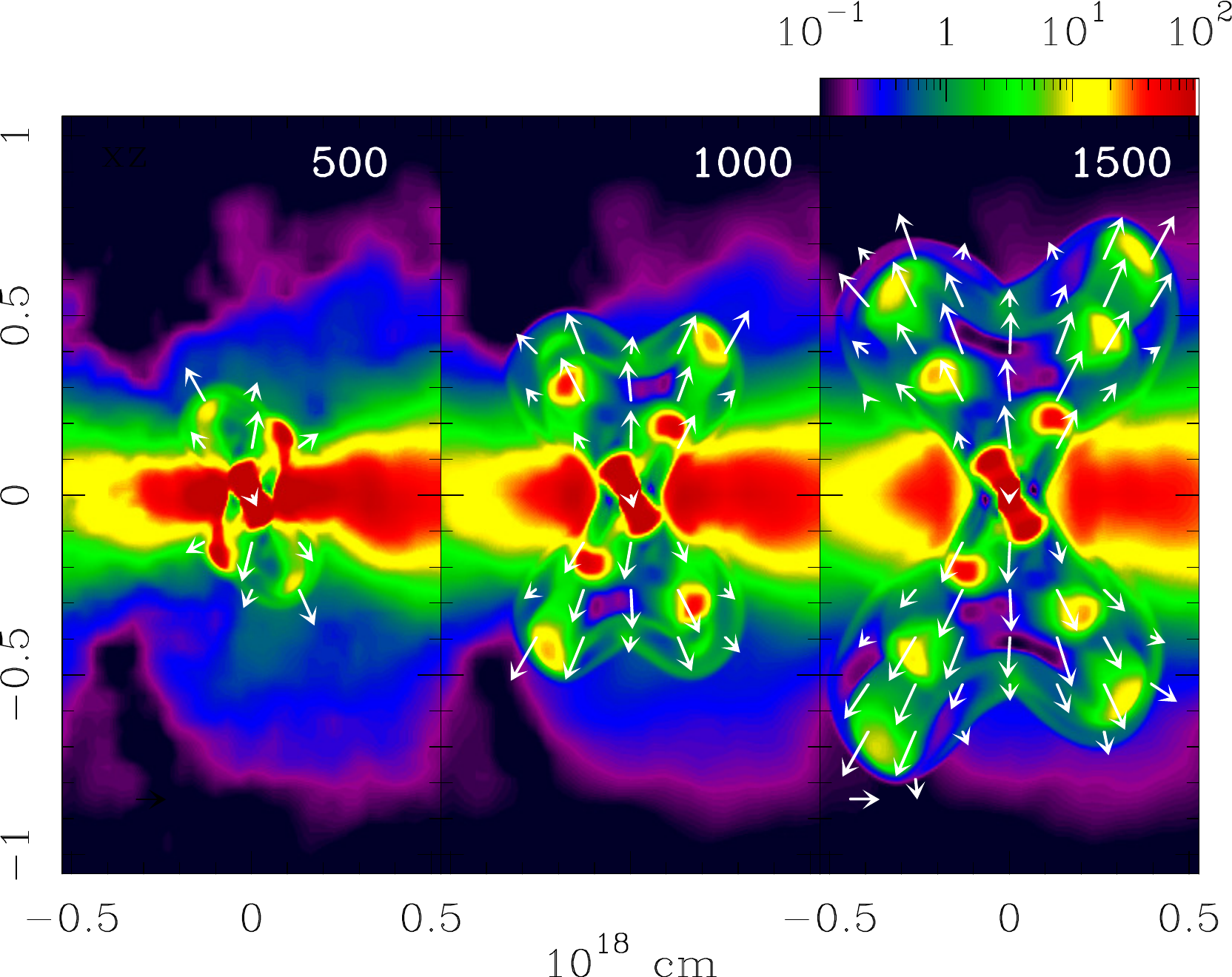}
 
  \includegraphics[width=1\linewidth]{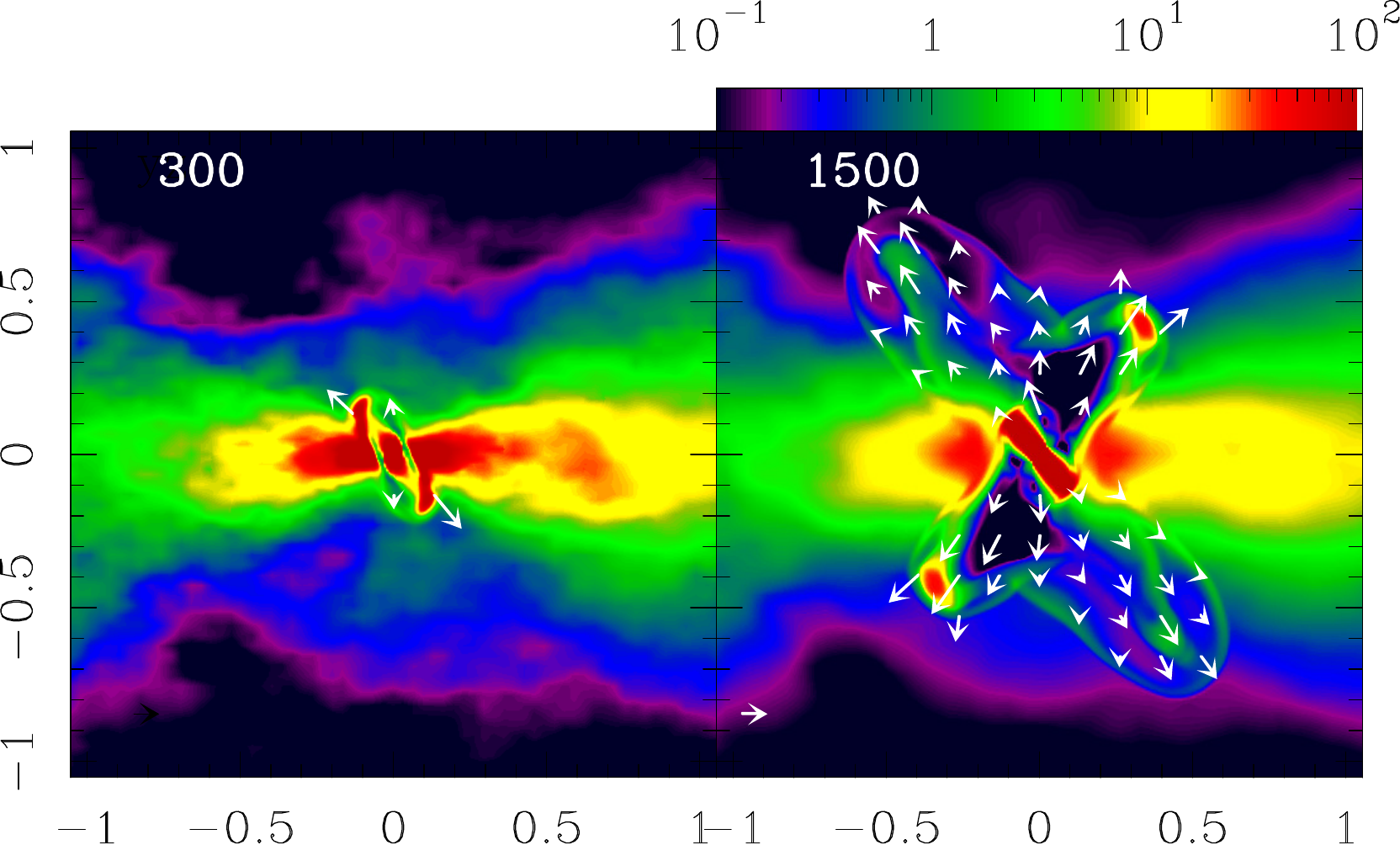}

 \caption{The same as Figure \ref{fig:velDenAlfa5cm} but for the $yz$-projection. At evolutiom time of 1500 yr, model 1 (top) shows six  ejections while model 2 (bottom) only show two ejections.}
  \label{fig:velDenAlfap30_cm}
\end{figure}

The temporal evolution of the electron density maps on the yz-projection for these models,  are shown in Fig. \ref{fig:velDenAlfap30_cm}. As in Fig. \ref{fig:velDenAlfa5cm}, the maps for integration times of 500, 1000, and 1500 yr are displayed for model 1. For model 2, due to the precession angle is larger, $\alpha$ = 40$^{\circ}$, we have enlarged the computational domain  in the y-direction for the jet to fit into the computational domain. Therefore, we only show two evolution times for this model of 300 and 1500 yr.

 For all density maps, both axes are given in units of 10$^{18}$ cm and the logarithmic color scale gives the number density in units of cm$^{-3}$.

In Figs. \ref{fig:velDenAlfa5cm} and \ref{fig:velDenAlfap30_cm} the noisy structure of
the AGB wind (a not uniform torus) produced by the modulation of the density through  a fractal structure, is evident. 

For both cases, in the xz-projection we observe a bipolar planetary nebula (see Fig. \ref{fig:velDenAlfa5cm}), while in the yz-projection  (Fig. \ref{fig:velDenAlfap30_cm}) a quadrupolar nebula is obtained \citep[see][for a discussion]{Velazquez2012}.

\subsection{Synthetic PV diagrams}
To compare with observed PV diagrams of point-symmetric planetary nebulae, from our hydrodynamic models we generate an array of synthetic PV diagrams at evolution time of 1500 yr, where we have considered different angles of inclination of the $z$-axis of the model with respect to the line of sight ($\phi$), as well as different counterclockwise rotations in the plane of the sky ($\psi$). Besides we have considered both xz- and yz-projections. 

\subsubsection{The xz-projection}
\begin{figure}
\centering
   \includegraphics[width=1.05\linewidth]{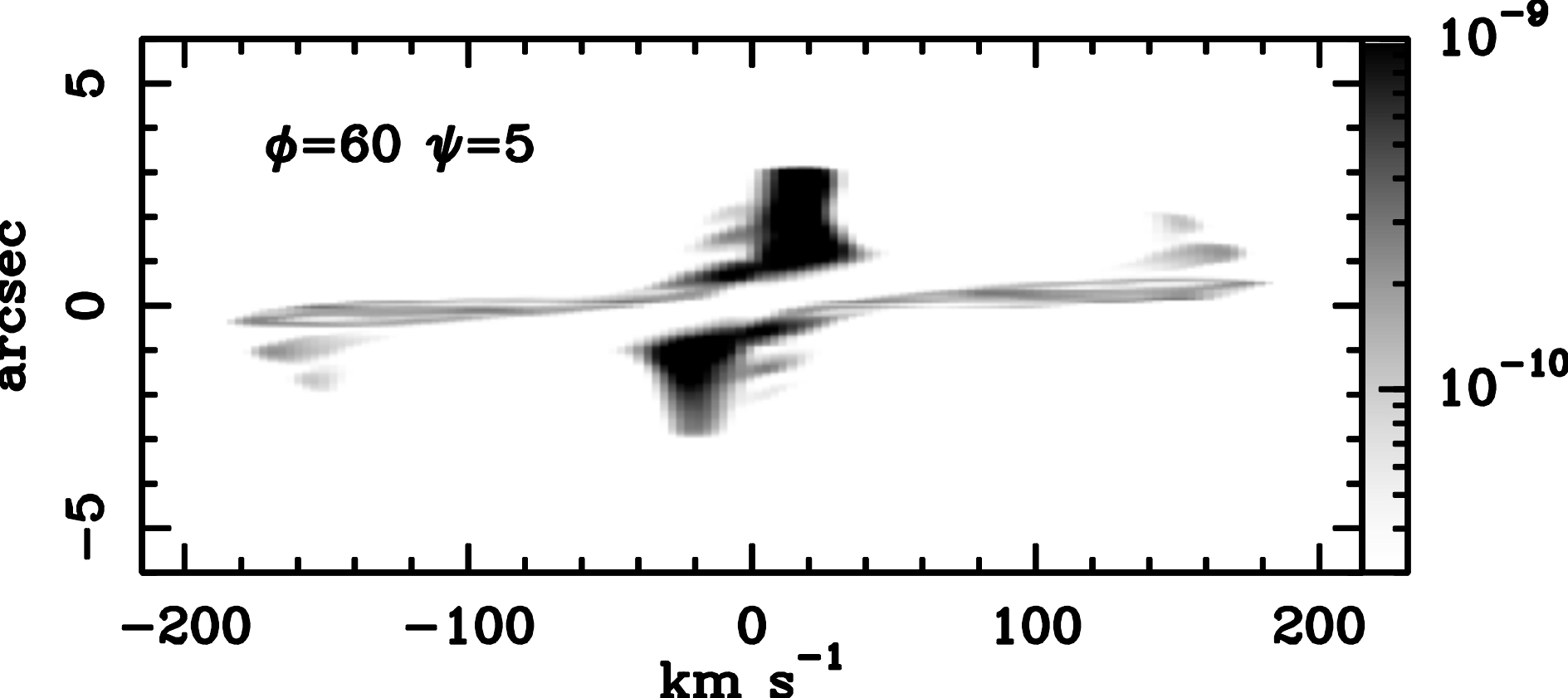}
   \includegraphics[width=1.05\linewidth]{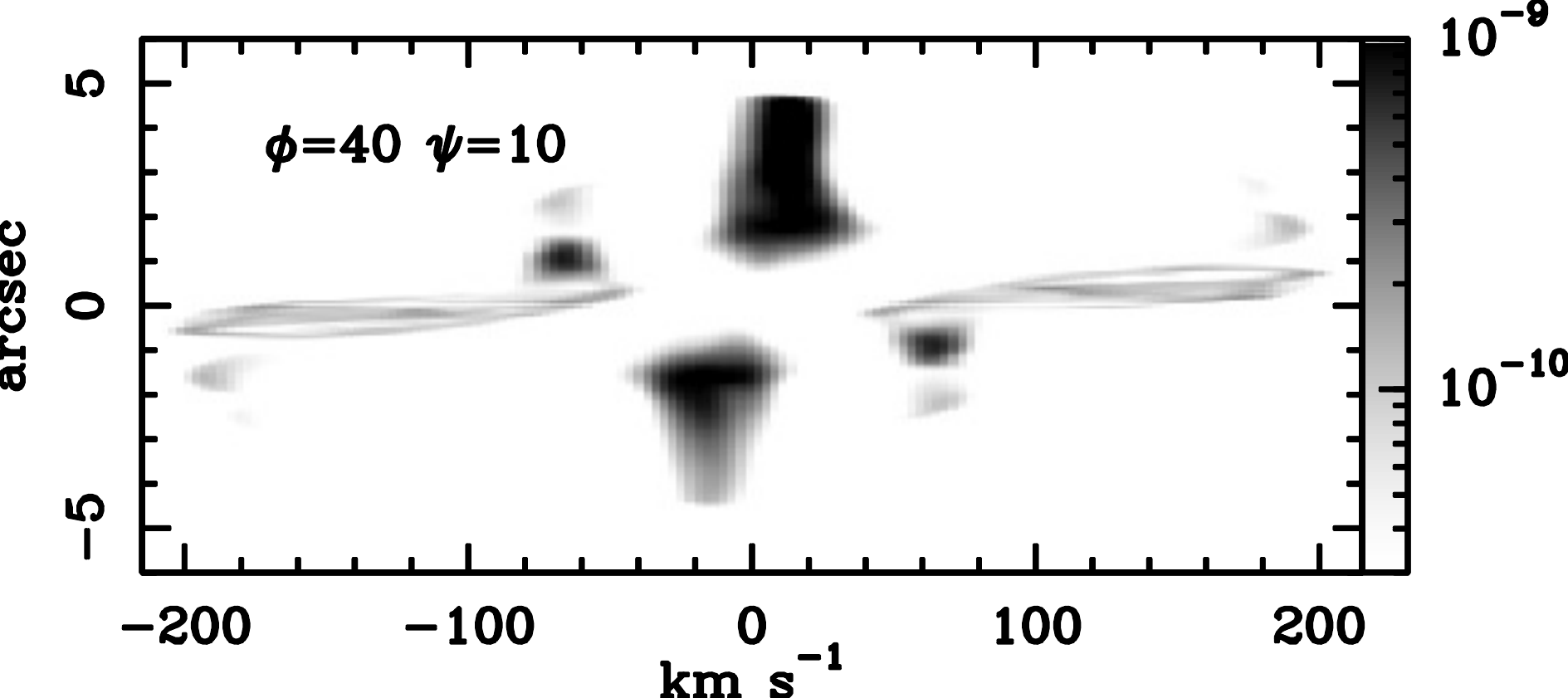}
   \includegraphics[width=1.05\linewidth]{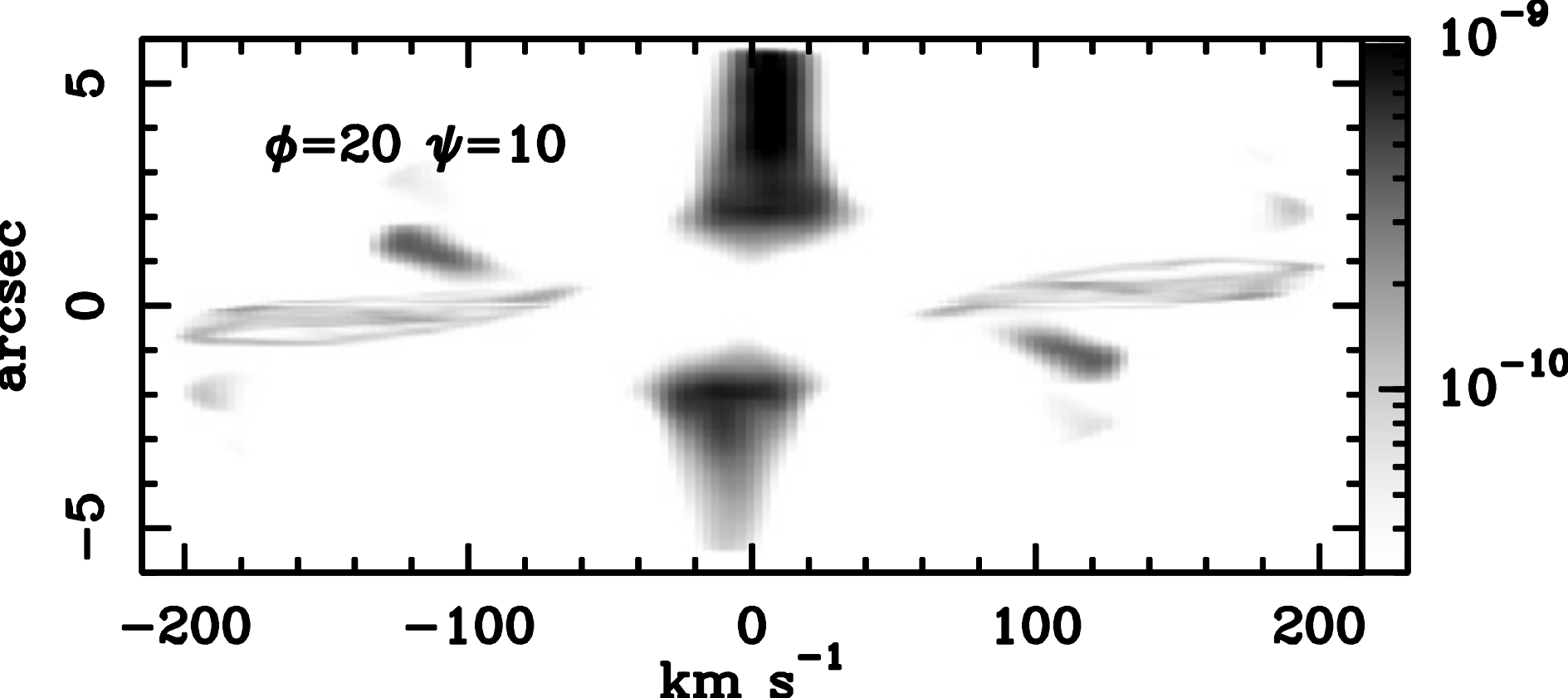}
 \caption{Synthetic [N\,{\sc ii}]$\lambda$6583 PV diagrams ($xz-$projection) for a slit
going through the centre of the source, corresponding to model 1 at evolution time 1500 yr.  These PV diagrams were obtained after applying two rotations to the computational domain: $\phi$ is the angle with respect to the line of sight, and $\psi$ is the angle with respect to the z-direction. The horizontal and vertical axes represent the radial velocity in km s$^{-1}$ and the spatial direction in arcsec (if we consider a distance to the source of 5.88 kpc), respectively.
The logarithmic grey-scale is given in units of erg cm$^{-3}$ sr$^{-1}$.}
  \label{fig:pvniixz13phi30Rot5CEN_new4nov_alfa30}
\end{figure}

\begin{figure}
\centering  
\includegraphics[width=1.0\linewidth]{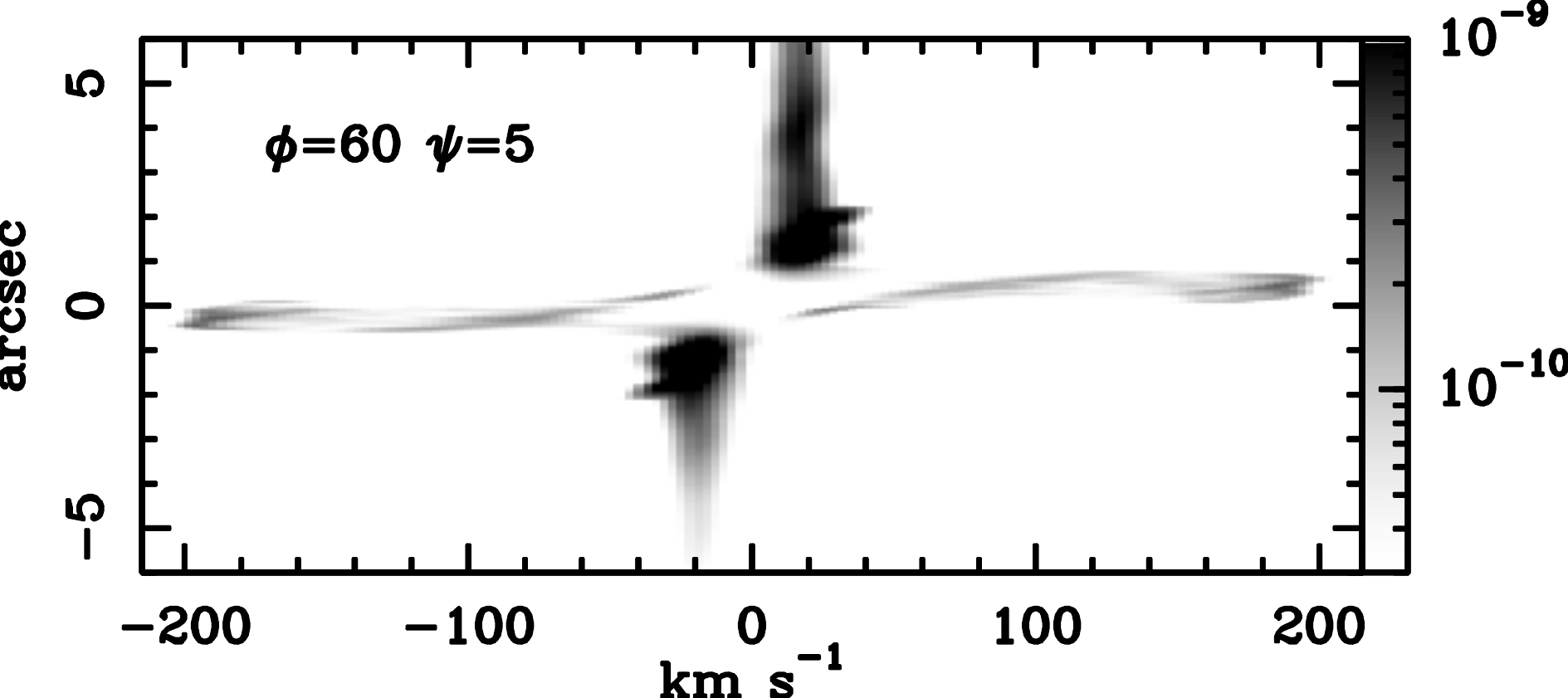}
\includegraphics[width=1.0\linewidth]{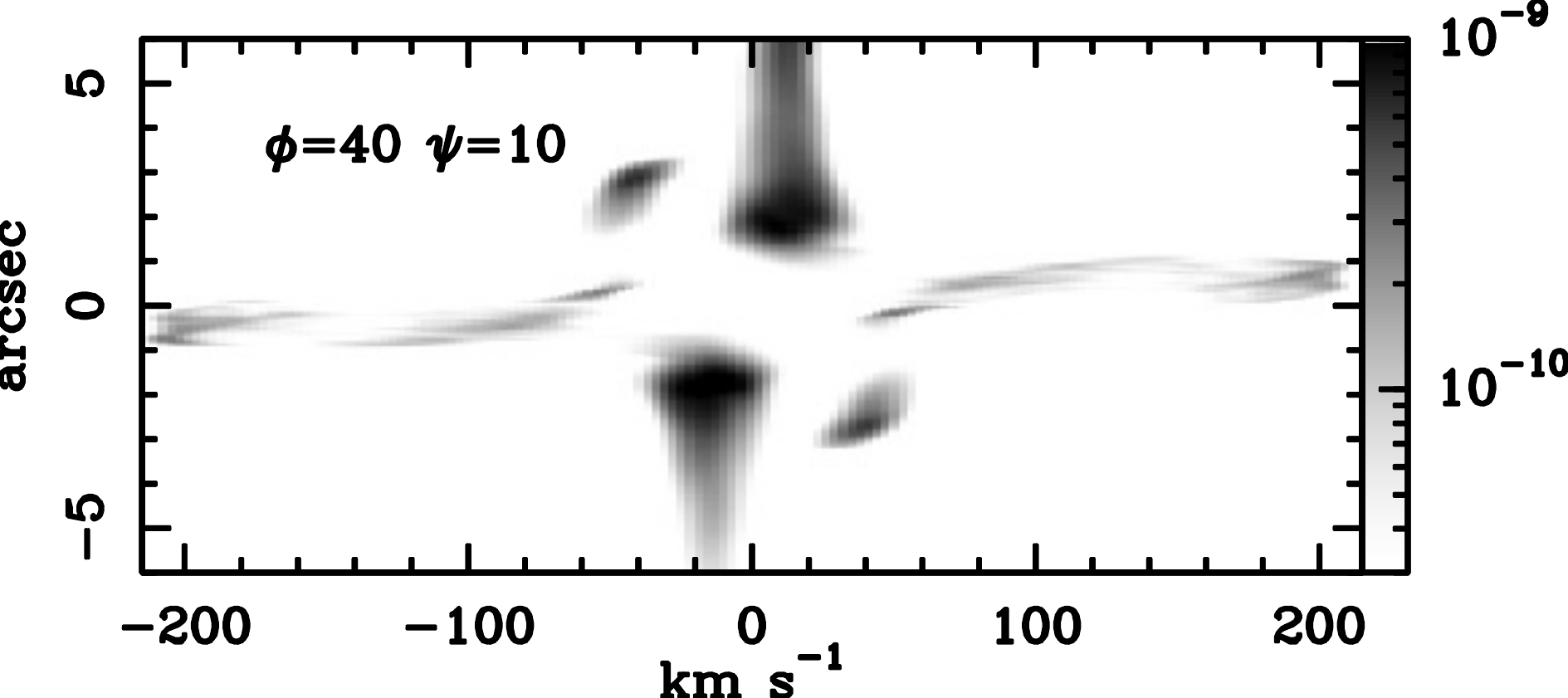}
\includegraphics[width=1.0\linewidth]{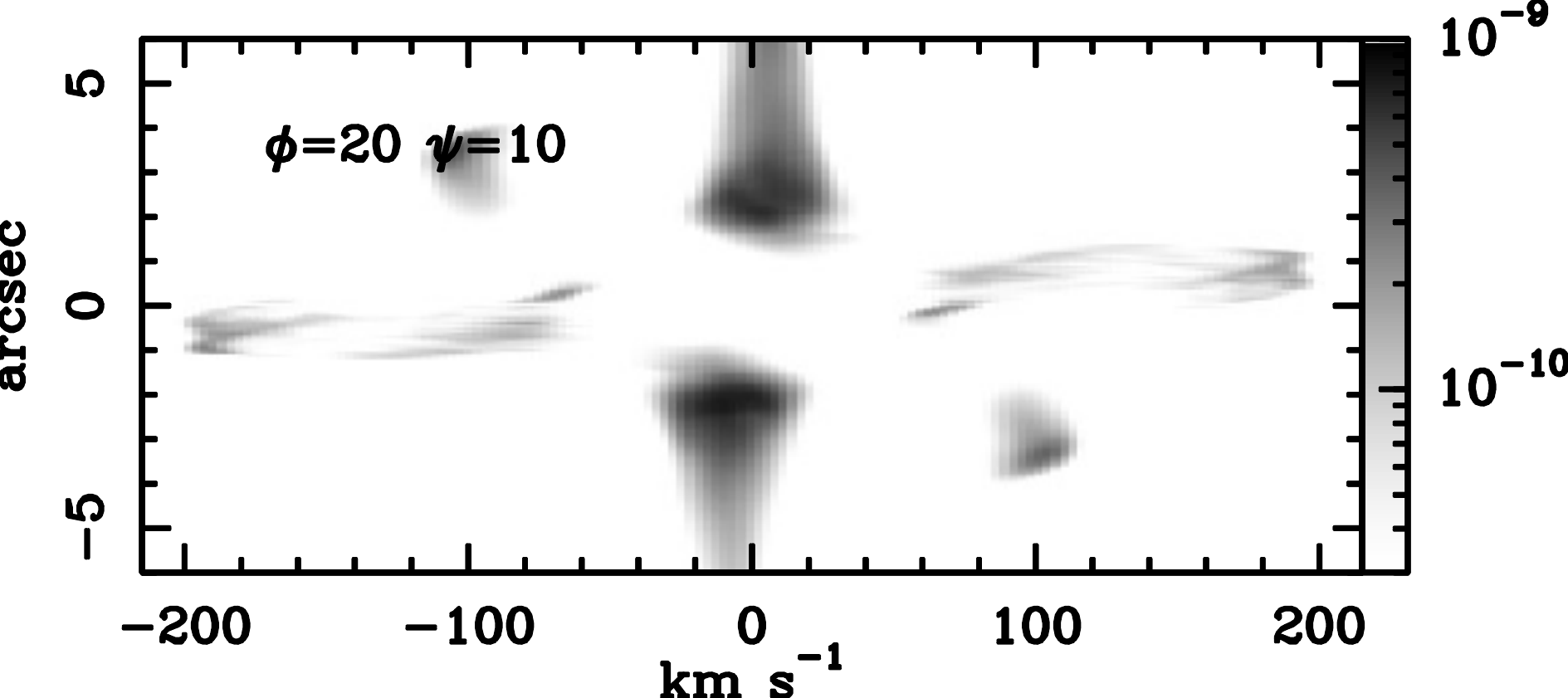}
 \caption{The same as Fig. \ref{fig:pvniixz13phi30Rot5CEN_new4nov_alfa30} but for model 2 and $xz-$projection.}
  \label{fig:pvniixz15phi30Rot5CEN_new4nov_alfa40}
\end{figure}

Figs. 
\ref{fig:pvniixz13phi30Rot5CEN_new4nov_alfa30}  and \ref{fig:pvniixz15phi30Rot5CEN_new4nov_alfa40} show the synthetic [N\,{\sc ii}] $\lambda$6583 PV diagrams ($xz$-projection) for models 1 y 2, respectively. Top, middle, and bottom panels of these figures display the PV diagrams for $(\phi , \psi)=(60^{\circ},5^{\circ});(40^{\circ},10^{\circ});(20^{\circ},10^{\circ})$.

 In Fig. \ref{fig:pvniixz13phi30Rot5CEN_new4nov_alfa30}, the PV diagrams  for model 1 show two bright condensations with velocities close to 0 km s$^{-1}$ for $\phi =20^{\circ}$, which correspond to the dense torus. As the angle $\phi$ increases these bright condensations appear slightly more separated in velocity (up to $\pm$20 km s$^{-1}$ for $\phi$ = 60$^{\circ}$). The structures associated with the ejections from the precessing jet, include a long filament at position 0 arcsec, which contains several ejections extending in velocity up to $\pm$200 km s$^{-1}$ and some faint knots, spatially separated from the centre by 2 or 3 arcsec
(see Fig. \ref{fig:pvniixz13phi30Rot5CEN_new4nov_alfa30}, bottom).
As the angle $\phi$ increases these knots show lower radial velocities, as a consequence of the projection effect (see Fig. \ref{fig:pvniixz13phi30Rot5CEN_new4nov_alfa30}, middle). In Fig. \ref{fig:pvniixz13phi30Rot5CEN_new4nov_alfa30} top, where $\phi=60^{\circ}$, the knots  are overlapped with the emission associated with the dense torus.\\

The PV diagrams for model 2 are shown in Fig. \ref{fig:pvniixz15phi30Rot5CEN_new4nov_alfa40}. The behaviour of the bright condensations (dense torus) is similar to model 1 and also the filament shows the same behaviour as in model 1. For this model, the precession angle is 40$^{\circ}$, consequently the two knots, corresponding to the  ejections, appear more spatially separated when $\phi=20^{\circ}$ (about 4 arc seconds from the centre, fig. \ref{fig:pvniixz15phi30Rot5CEN_new4nov_alfa40} bottom). As $\phi$ increases these features show lower velocity and appear closer to the centre. When $\phi = 60^{\circ}$ both features are overlapped with the torus, similar to model 1. 

\subsubsection{The yz-proyection}
\begin{figure}
\centering
   \includegraphics[width=1.05\linewidth]{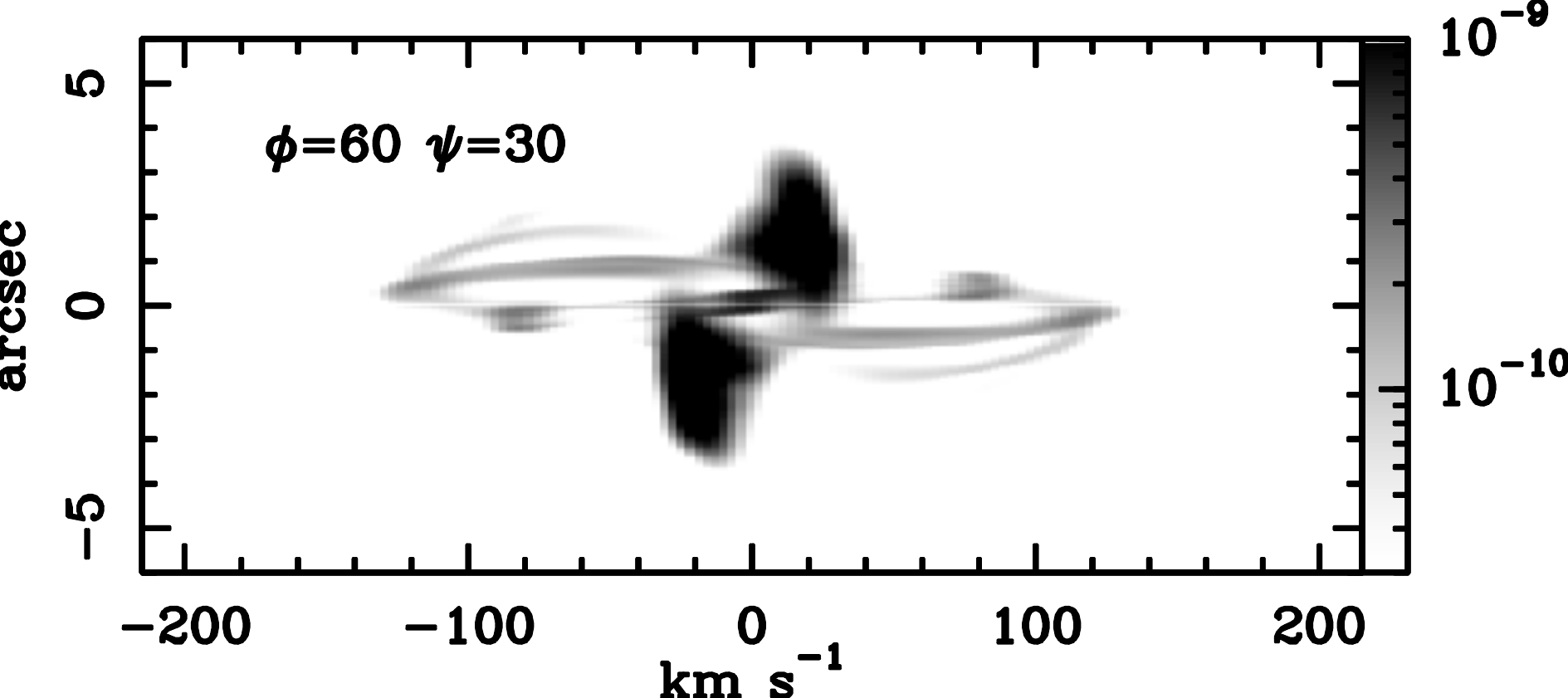}
   \includegraphics[width=1.05\linewidth]{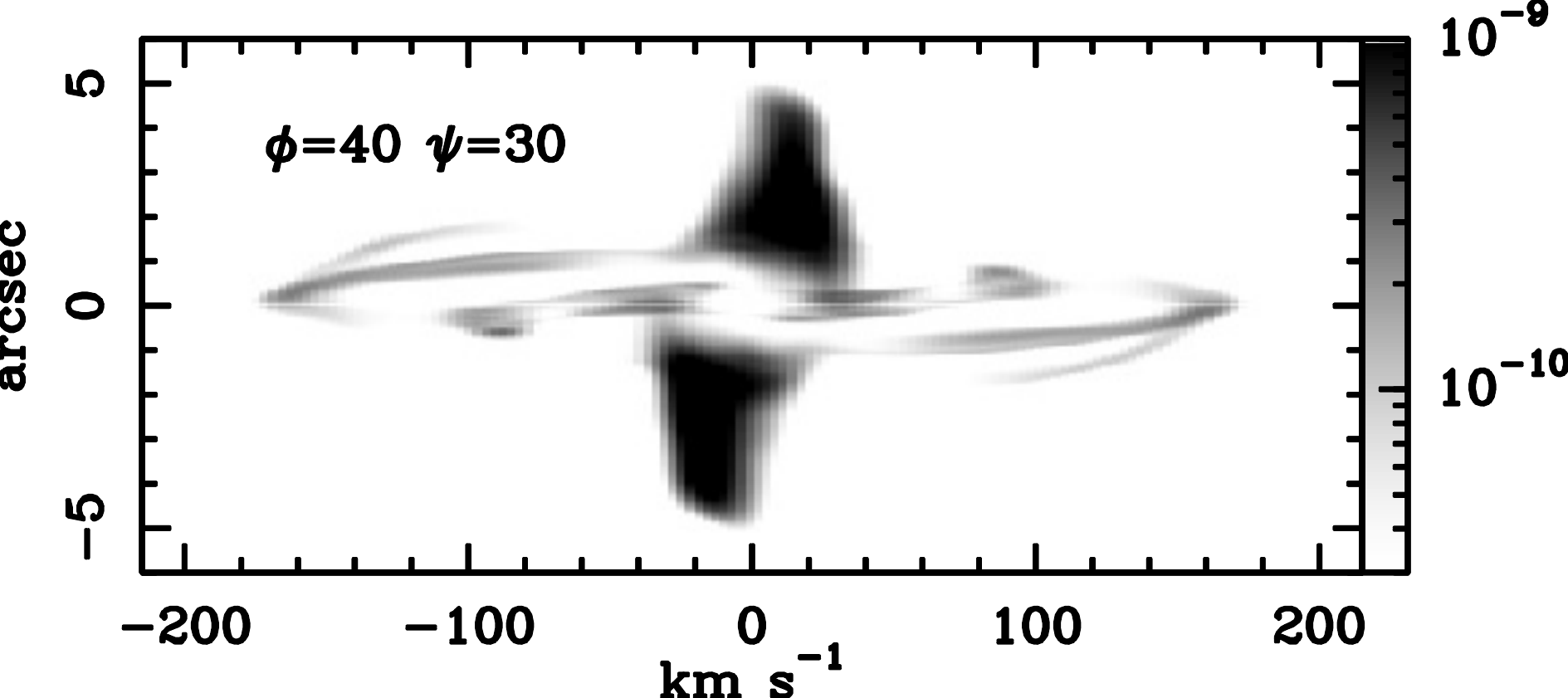}
   \includegraphics[width=1.05\linewidth]{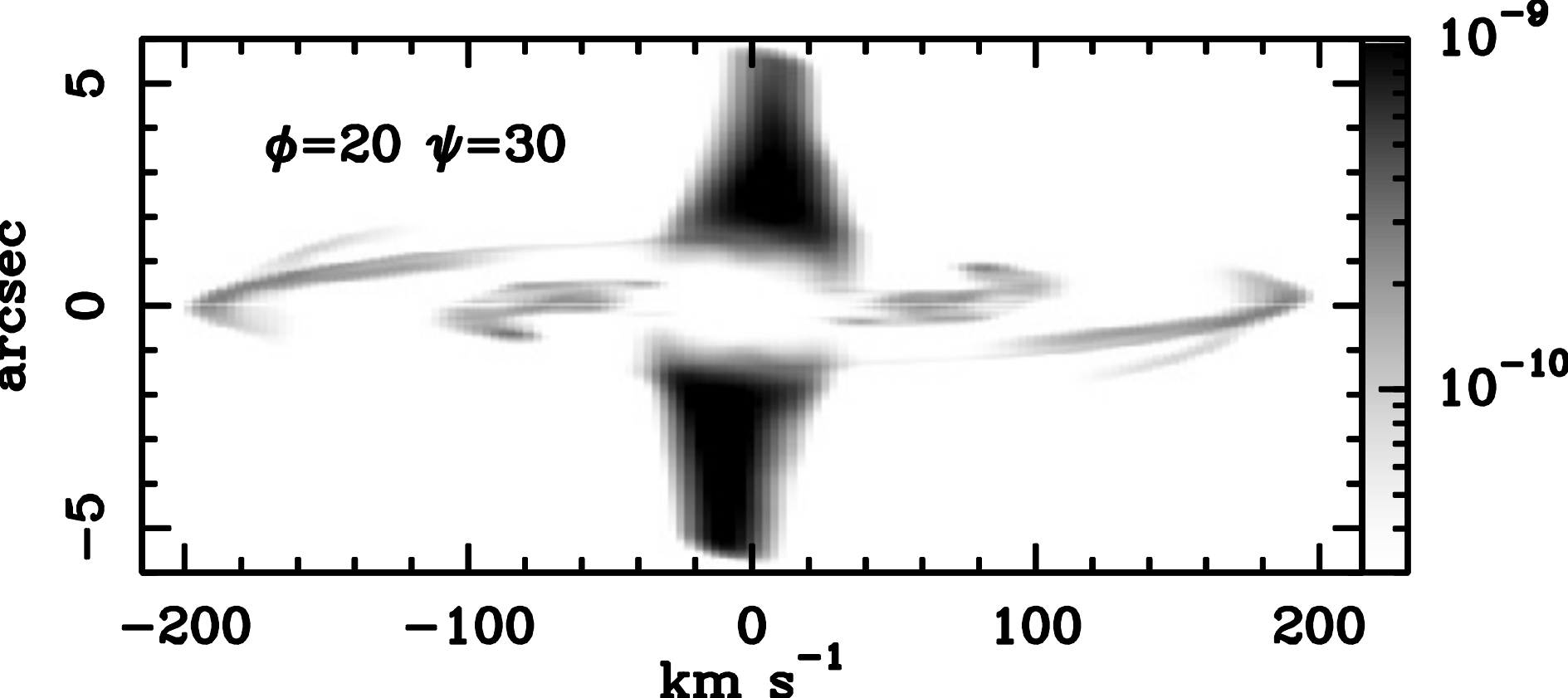}  
 \caption{The same as Figure \ref{fig:pvniixz13phi30Rot5CEN_new4nov_alfa30} for model 1 and $yz-$projection.}
  \label{fig:pvniiyz13phi30Rot30}
\end{figure}

\begin{figure}
\centering
   \includegraphics[width=1.05\linewidth]{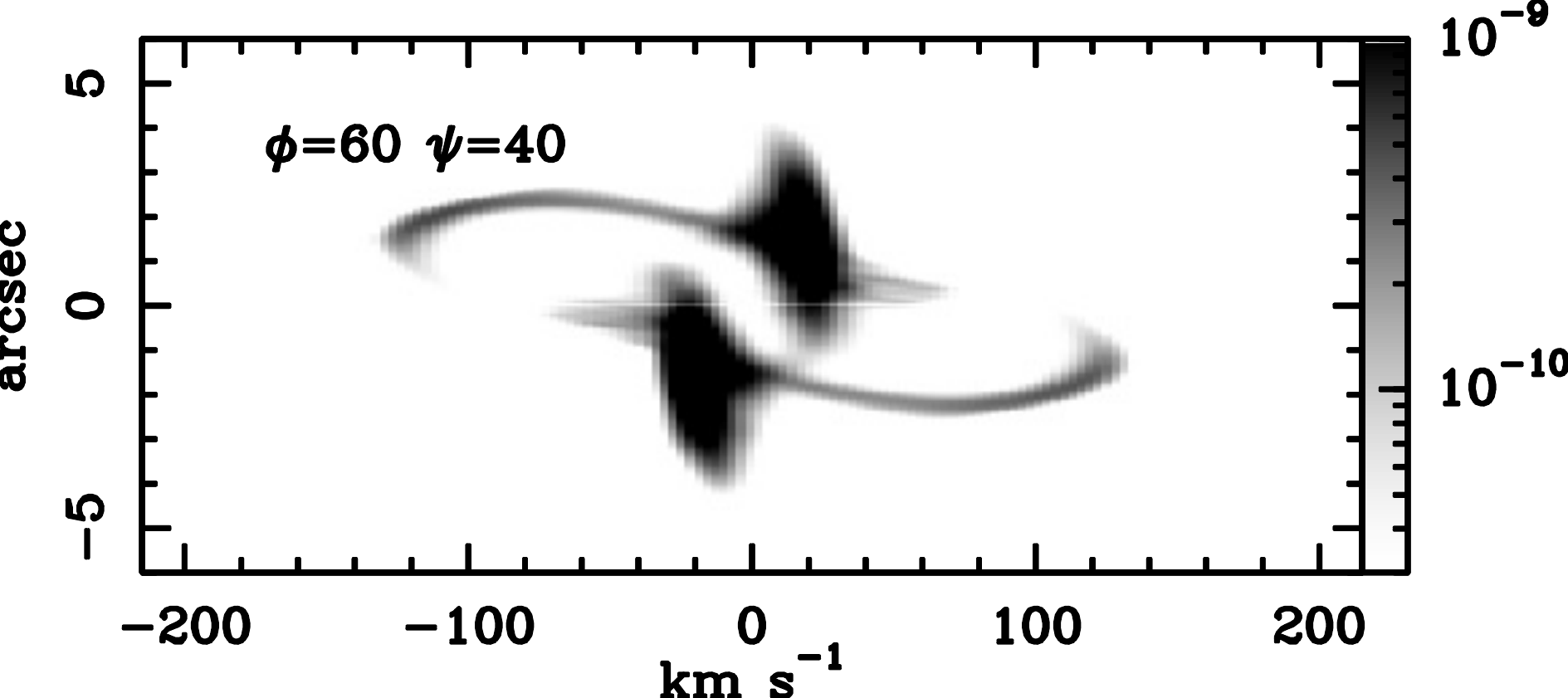}
   \includegraphics[width=1.05\linewidth]{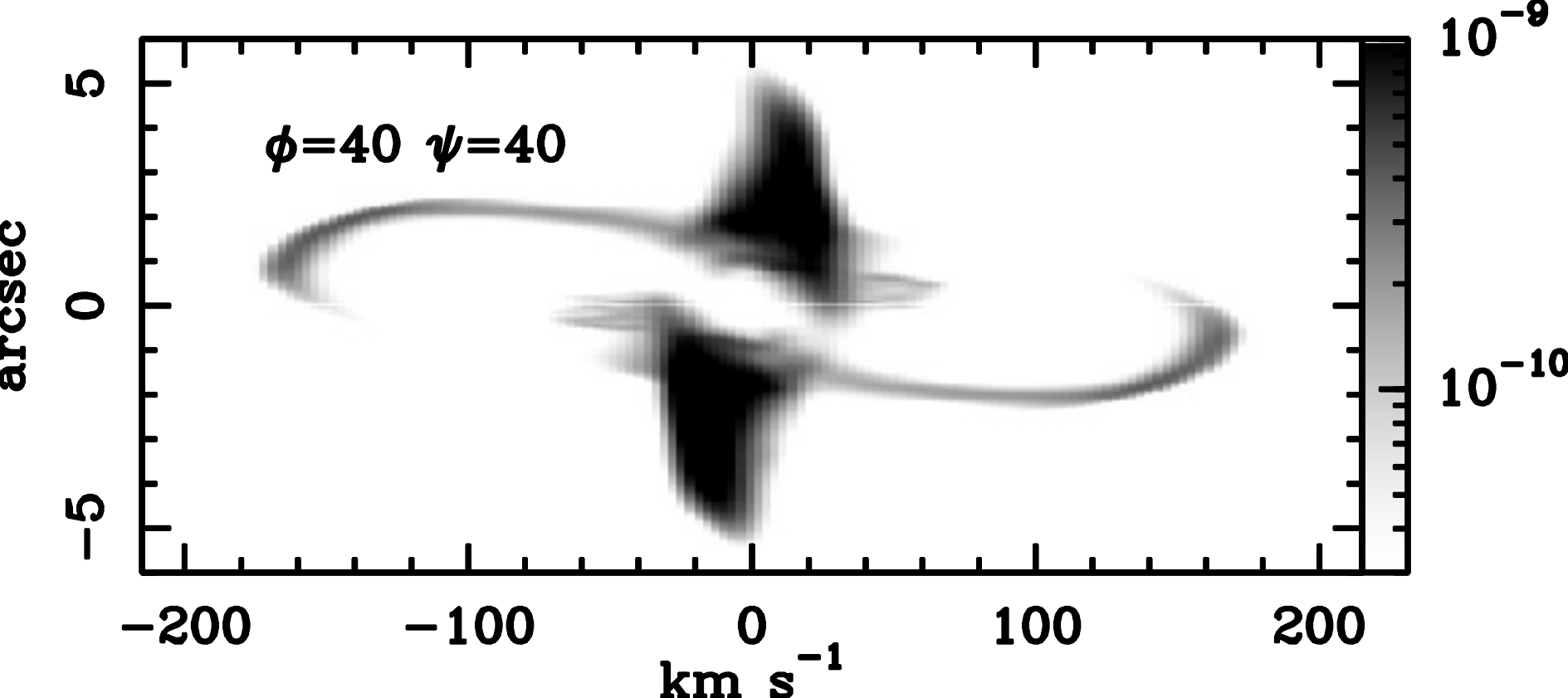}
   \includegraphics[width=1.05\linewidth]{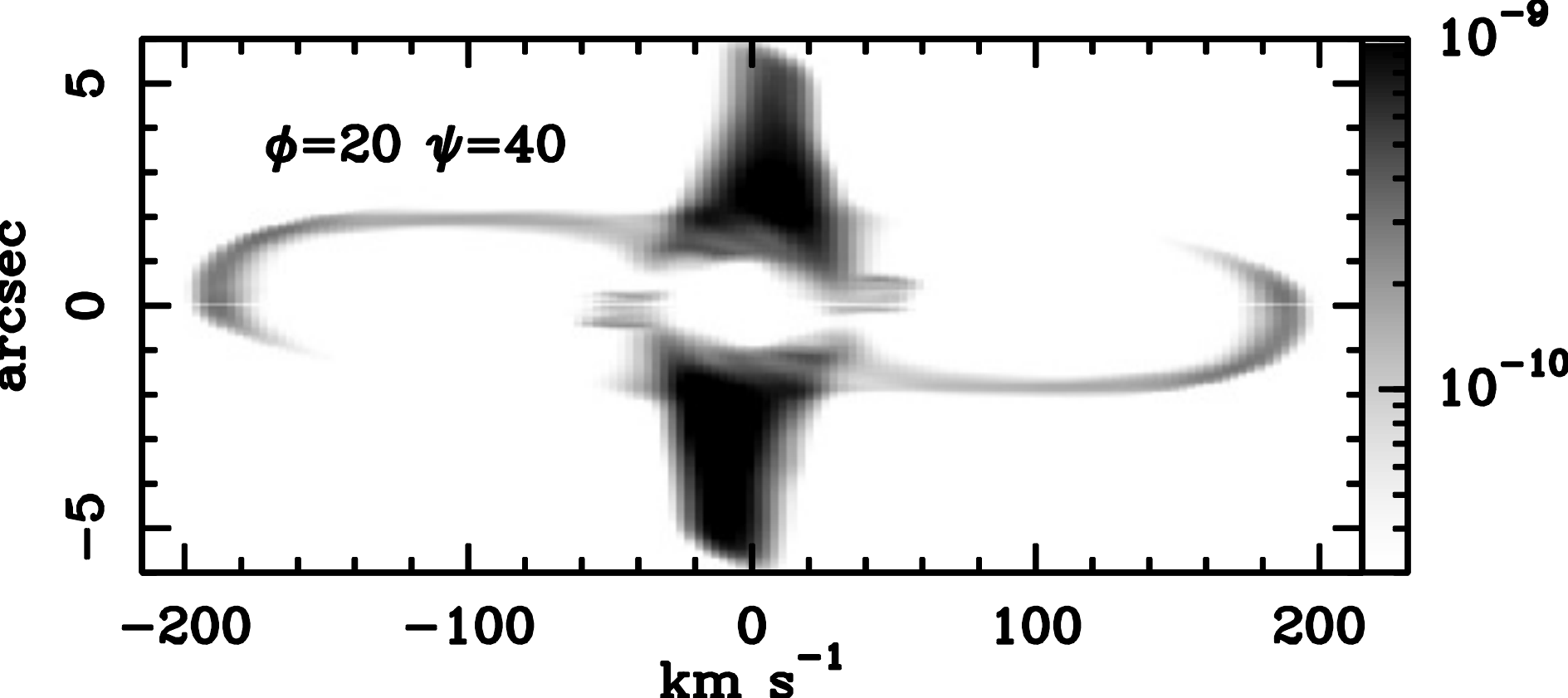}
   
 \caption{The same as Figure \ref{fig:pvniixz13phi30Rot5CEN_new4nov_alfa30} but for model 2 and $yz-$projection. } 
  \label{fig:pvniiyz15phi40Rot40}
\end{figure}

The PV diagrams for models 1 and 2 obtained for the $yz$-projection are shown in Figs. 
\ref{fig:pvniiyz13phi30Rot30} and \ref{fig:pvniiyz15phi40Rot40}, respectively. Both PV diagrams  display similar morphologies. Several filaments are observed along the horizontal axis, which are produced by the variability of the jet ejections. All the  filaments display a S-like shape. Model 1 has more filaments due it has a factor of precession of $f_p$=4, while model 2 has $f_p$=1.5. The filaments result more separated along the position axis for model 2 than for model 1. This is a consequence of a larger semi-aperture angle of precession (40$^{\circ}$ and 30$^{\circ}$, respectively).

\section{The S-shape spectrum of H1-67: Observational Results}

As it was said in the Introduction the PN H\,1-67 presents S-shape spectra as obtained with the MES at OAN-SPM with the slit is located along its symmetry axis. 
Therefore this is a suitable object to compare its PV diagrams with our numerical simulations. 
Some characteristics of this and other similar objects are listed in Table 2.

H\,1-67 is located towards the galactic bulge, showing a heliocentric radial velocity of $-$8.05 km s$^{-1}$
(this work). 
Its heliocentric distance is 5.88
 kpc \citep{FrewParker2016} and it is situated at a height of $-$451.6 pc (this  work). It is ionised by a [WO\,2] central star \citep{AckerNeiner2003}. 
The total nebular abundance ratios for this PN are
 N/O = 0.56 and He/H = 0.125 \citep{EscuderoCosta2004}, therefore, this object is a
 nitrogen and helium rich PN (Peimbert Type I PN). Morphologically it has been classified as a bipolar 
 nebula \citep{ReesZijlstra2013}. From our spectroscopic analysis it is found that the nebular
 morphology of H\,1-67 is highly asymmetric, showing the appearance of broken face-on torus (Fig. \ref{fig:H167image})
with low expansion velocity, and also knots and
 jets at high velocities of $\pm$100 km s$^{-1}$. 
 
In this work, we use high-resolution spectra to analyse the kinematical
behaviour of the gas in this [WO 2]PN and, by comparing it with the
hydrodynamical models presented above, we investigate if the morphological structure of
H\,1-67 can be produced by the action a precessing jet
and a dense torus. We compare the
position-velocity diagrams obtained from the observations with those
predicted by the hydrodynamical models.
\\

\begin{table*}
\centering
 \begin{minipage}{100mm}
  \caption{Main characteristics of H\,1-67 and other PNe}
\begin{tabular}{lllllllll}
\hline \hline 
PN G & Name  & Stellar & $V_{hel}$&Dmean&z&N$^a$ & O$^a$\\
& &Type& km s$^{-1}$&kpc&pc&\\
\hline
009.8$-$04.6&H\,1-67&[WO\,2]&  $-$8.05&5.88$\pm$1.76& $-$451.6 &8.50& 8.75\\ 
003.2$-$06.2&M\,2-36&possible binary& +57.0&6.99&--&8.42& 8.85\\
353.7$-$12.8&Wray\,16-411&--&$-$50.0&6.14$\pm$1.75& --&--& ---\\
\hline
\multicolumn{5}{l}{$^a$ In 12 + log $X$/H}  (see \citet{EscuderoCosta2004} and \citet{LiuLuo2001})
\end{tabular}
\label{tab:charac}
\end{minipage}
\end{table*}

\subsection{Long-Slit Spectroscopy}

The Manchester Echelle Spectrometer \citep[MES,][]{Meaburn2003} is a long-slit echelle spectrometer, which 
uses narrow-band filters to isolate the orders containing the emission lines. The 90 \AA ~ bandwidth filter allowed us to obtain the H$\alpha$ and [N\,{\sc ii}] nebular
emission  lines. In this paper, we only show the behavior of the
[N\,{\sc ii}] $\lambda$6583 line because H$\alpha$ shows the same structure.

The Marconi 2 CCD was used as detector, which has a pixel size of 13.5 $\mu$m. A binning of 2$\times$2 pixel was applied in the observations and, using the secondary mirror f/7.5 which provides a plate scale of 13.2 $''$/mm, the plate  scale on the  detector was 0.356 $''$/pixel. The  slit width 
used was 150 $\micron$ ($2''$ on the sky), providing a spectral resolution of 11 km s$^{-1}$.

We obtained observations from two slit positions over H\,1-67, one position with the slit at P.A. = 45$^{\circ}$ across the centre of the nebula corresponds to its symmetry axis.
The other position with the slit at P.A.= 0$^{\circ}$, lies on the East side. The slit positions are indicated
in Fig. \ref{fig:H1-67pv} (left side) on H$\alpha$ images obtained by us. 
The integration time for the spectra was 900 s for both slit positions. 
 Immediately after every science observation, 
exposures of a Th-Ar lamp were acquired for wavelength calibration.  The internal precision 
of the  lamp calibrations
 is better than 1.0 km s$^{-1}$. The spectra were reduced using the standard
  processes for the MES data, using IRAF\footnote{IRAF is distributed by the National Optical Astronomy Observatories, which is operated by the Association of Universities for Research in Astronomy, Inc., under contract to the National Science Foundation.} tasks to correct for bias and calibrate in wavelength.

\subsection{Position-Velocity diagrams of H\,1-67}
 \label{sec:space}
PV diagrams for this PN were obtained with the WIP
software \citep{Morgan1995}. In Fig. \ref{fig:H1-67pv} right side, the PV diagrams 
are shown for the slits located in the two positions, P.A. = 45$^{\circ}$ and P.A. = 0$^{\circ}$.
In the horizontal axis the heliocentric
radial velocity is represented and the vertical axis shows the spatial
direction. The emission line shown is [N\,{\sc
  ii}] $\lambda$6583.\\

 In the PV diagram with the slit at P.A. = 45$^{\circ}$, two bright inclined condensations (at a heliocentric systemic velocity of $-$8.05  km s$^{-1}$) are visible, they are labeled as B and C in Fig. \ref{fig:H1-67pv} (top). Two knots at high velocity  of $\pm$97.07 km s$^{-1}$ (relative to the systemic velocity) are seen,  labeled as A and D in Fig. \ref{fig:H1-67pv} (top).
 In the PV diagram with the slit at P.A. = 0$^{\circ}$, two bright condensations are visible (they are labeled as B' and C' in Fig. \ref{fig:H1-67pv}, bottom) and two knots at high velocity of $\pm$96.4 km s$^{-1}$ (relative to the systemic velocity) are seen, labeled as A' and D' in Fig. \ref{fig:H1-67pv}.
In addition, comparing both PV diagrams, we observe that the spectra obtained with the slit at P.A. =45$^{\circ}$ shows a S-shape, differently from what is obtained with the slit at P.A. = 0$^{\circ}$. This suggests a bipolar, collimated jet whose ejection axis has changed following a precessing movement. 
The observations are not calibrated in flux, but we have added a scale in counts.
In the observational PV diagrams, the ratio between the brightest (torus) and faintest (filament)
regions is of the order of 20.

\subsubsection{Emission line profile}
\begin{figure}
  \begin{center}

\includegraphics[width=1.1\linewidth]{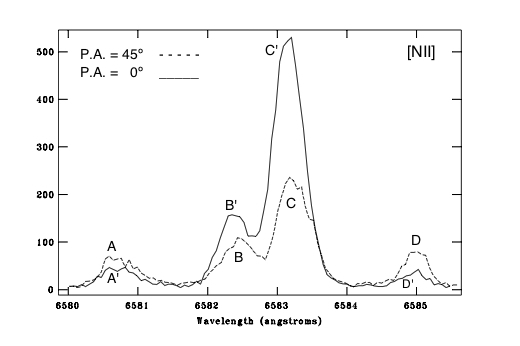}
  \caption{ Emission line profiles of H\,1-67 of the [N\,{\sc
        ii}] $\lambda$6583 line,  for two positions of the
    slit, P.A. = 45$^{\circ}$ (dashed line) and P.A. = 0$^{\circ}$ (solid line). Labels marked correspond
    to the bright condensations and a bipolar precessing jet ending in knots of the Fig. 
    \ref{fig:H1-67pv}.}
  \label{fig:H1-67profile}
 \end{center}
\end{figure}

One-dimensional profile of the [N\,{\sc  ii}] emission-line for the two positions of the slit are displayed in Fig. \ref{fig:H1-67profile}. The wavelength is represented in the horizontal axis, and the vertical axis shows the intensity in counts. The features marked in  Fig. \ref{fig:H1-67pv} (for both positions) are shown in this figure. Both profiles are kinematically similar.
We can distinguish two regions in this PN, an inner zone (B, C and B', C') and an outer zone (A, D and A', D'). 
The inner zone that corresponds to the clumpy torus is expanding at about 18 km s$^{-1}$ while the outer zone corresponding to a precessing jet ending in knots, shows an expansion velocity of about 97 km s$^{-1}$. 
Note that the radial heliocentric velocity in the knots marked as A,D and 
A',D' in both positions slits  are similar.
In Table \ref{tab:kinematics} we present the kinematics for each feature.

  \begin{table}
\centering
 \begin{minipage}{140mm}
   \caption{Kinematics of the features}
\begin{tabular}{lccccl}
 P.A. = 45$^{\circ}$\\
\hline \hline
Knot & A  & D &\vline &B & C \\
 & km s$^{-1}$&km s$^{-1}$&\vline & km s$^{-1}$&km s$^{-1}$\\
\hline
Vhelio&-103.7&90.40 &\vline &-26.3&10.6\\ 
Vexp&&97.06& \vline &&18.46\\ 
\hline
P.A. = 0$^{\circ}$\\
\hline \hline
Knot & A'  & D' &\vline &B' & C' \\
 & km s$^{-1}$&km s$^{-1}$&\vline & km s$^{-1}$&km s$^{-1}$\\
\hline
Vhelio&-103.4&89.3&\vline &-28.6&7.3\\ 
Vexp&&96.4& \vline &&18.0\\ 
\hline
\end{tabular}
\label{tab:kinematics}
\end{minipage}
\end{table}

\begin{figure}
\centering
   \includegraphics[width=1.0\linewidth]{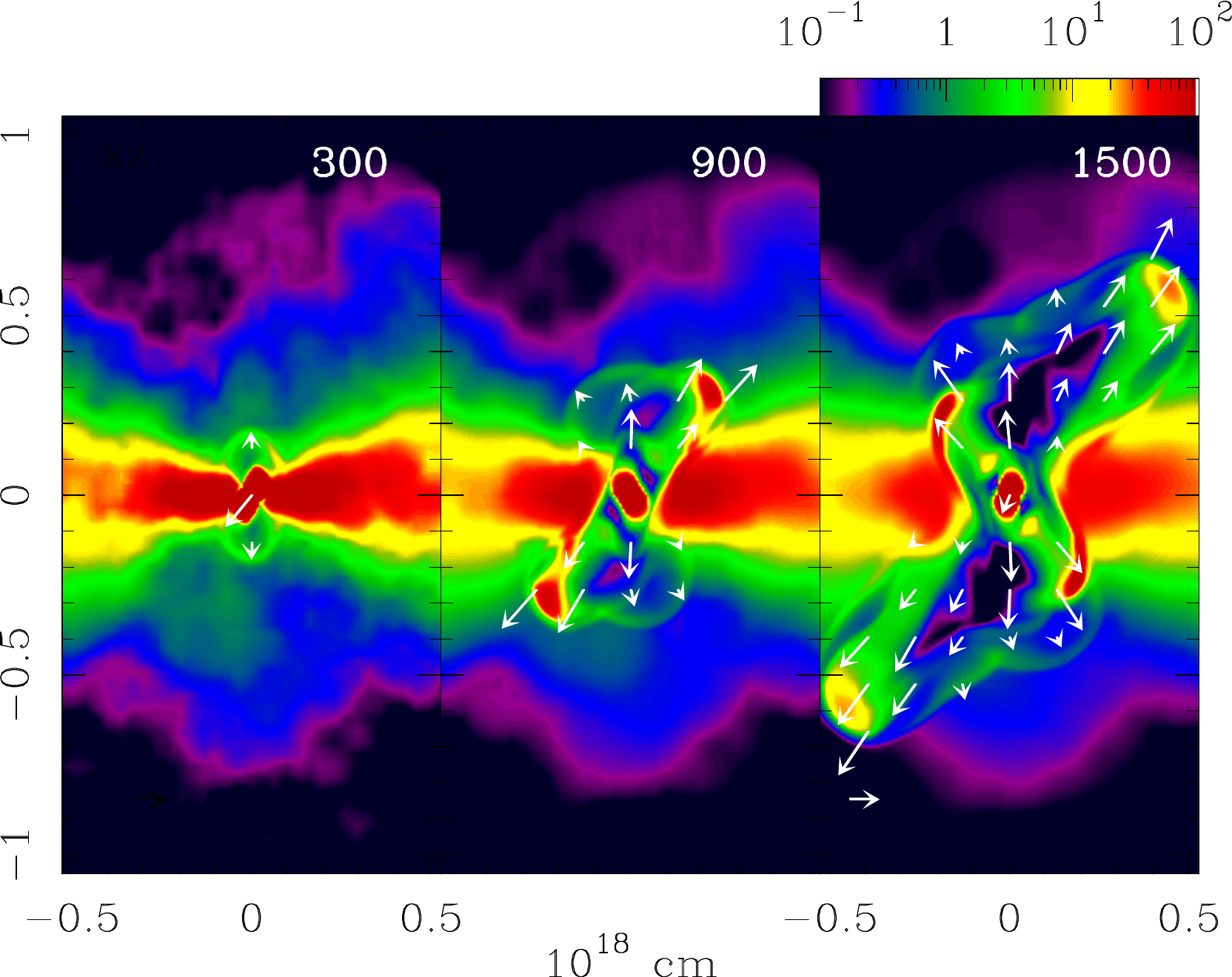}
   
   \includegraphics[width=1.0\linewidth]{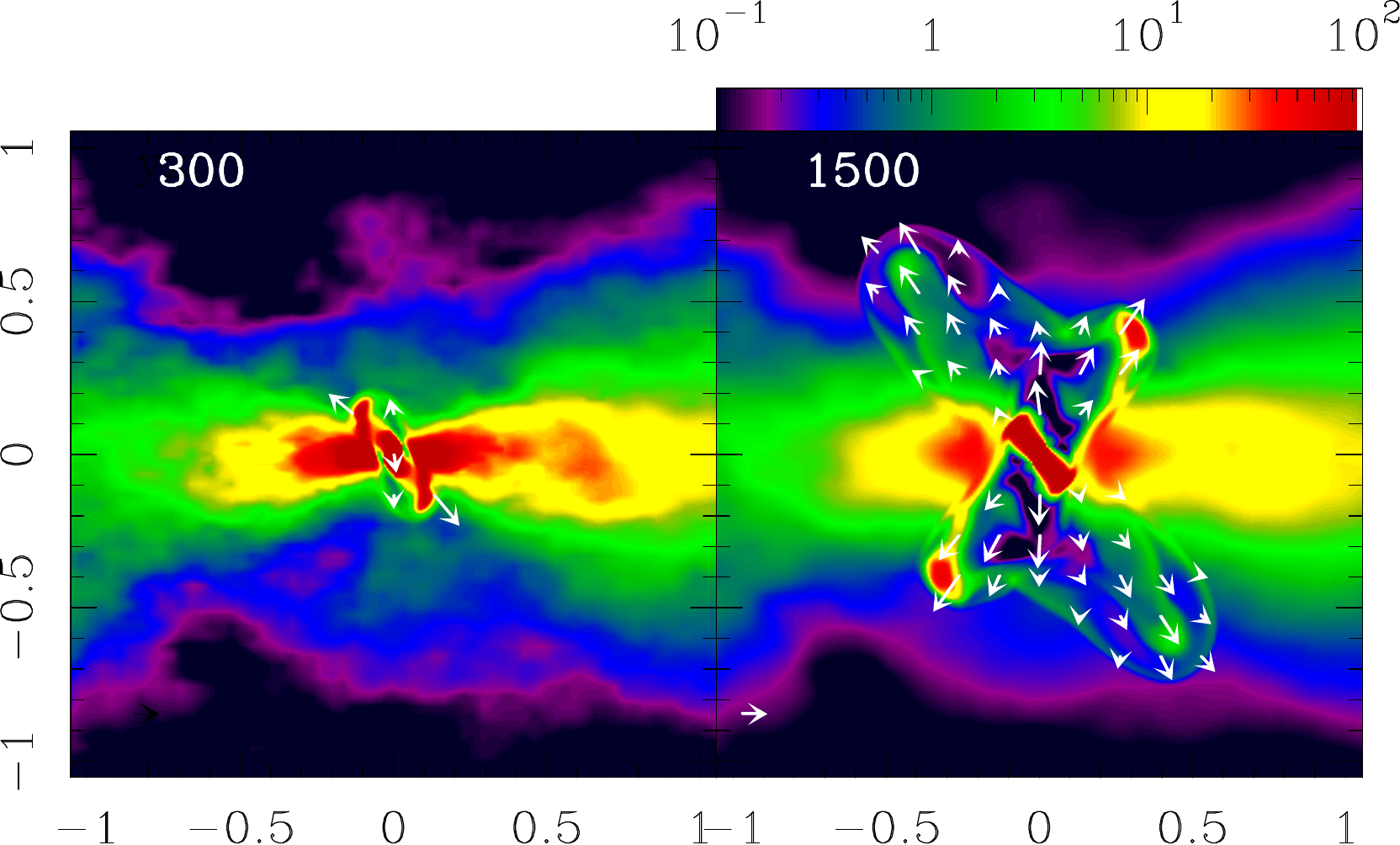}
 \caption{Temporal evolution of the electron density obtained for model 3. Upper panels show the evolution for the $xz$-projection while
 bottom panels display the evolution for the $yz$-projection.}
   \label{fig:velDenAlfap40_cmmodel3}
\end{figure}

\subsection{Computing models to reproduce H\,1-67 observational results: Models 3 and 4}

Analysing the synthetic PV diagrams obtained for the different models, we found  that the S-shape is reached in the yz-projections when the precession angle $\alpha$ is larger than 30$^{\circ}$. Whenever the angle is greater, the jet ejections open giving rise to a more defined S-shape. Also, the number of knots and filaments obtained depends on the number of precessions, $f_p$,  and the number of ejections in each turn, $f_q$.  Therefore, $f_p$ should be small to obtain less ejections in the PV diagrams (compare Fig. \ref{fig:pvniiyz13phi30Rot30} and Fig. \ref{fig:pvniiyz15phi40Rot40}).\\

We notice that the observed PV diagrams shown in Fig. \ref{fig:H1-67pv} for H\,1-67, have some features similar to those obtained in the PV diagrams of model 2, for the yz-projection. 
Due to this, we have carried out a 3rd model in order  to reproduce better the main characteristics of the observed PV diagrams of H\,1-67. In Model 3, which has $\alpha$=40$^{\circ}$ and $f_p$=1.5, the ratio between the precession period and the velocity variability period, factor $f_q$,  was set to 4.  Therefore model 3 is similar to model 2, but with more ejections per precession turn.\\

The electron density distribution map for this model in the xz-projection at evolution times of 300, 900, and 1500 yr, is presented in Fig. \ref{fig:velDenAlfap40_cmmodel3} (top) and the yz-projection is presented in Fig. \ref{fig:velDenAlfap40_cmmodel3}  (bottom) at evolution times of 300 and 1500 yr. As expected, these maps are similar to the ones for model 2.\\

 \begin{figure}
\centering  
\includegraphics[width=1.0\linewidth]{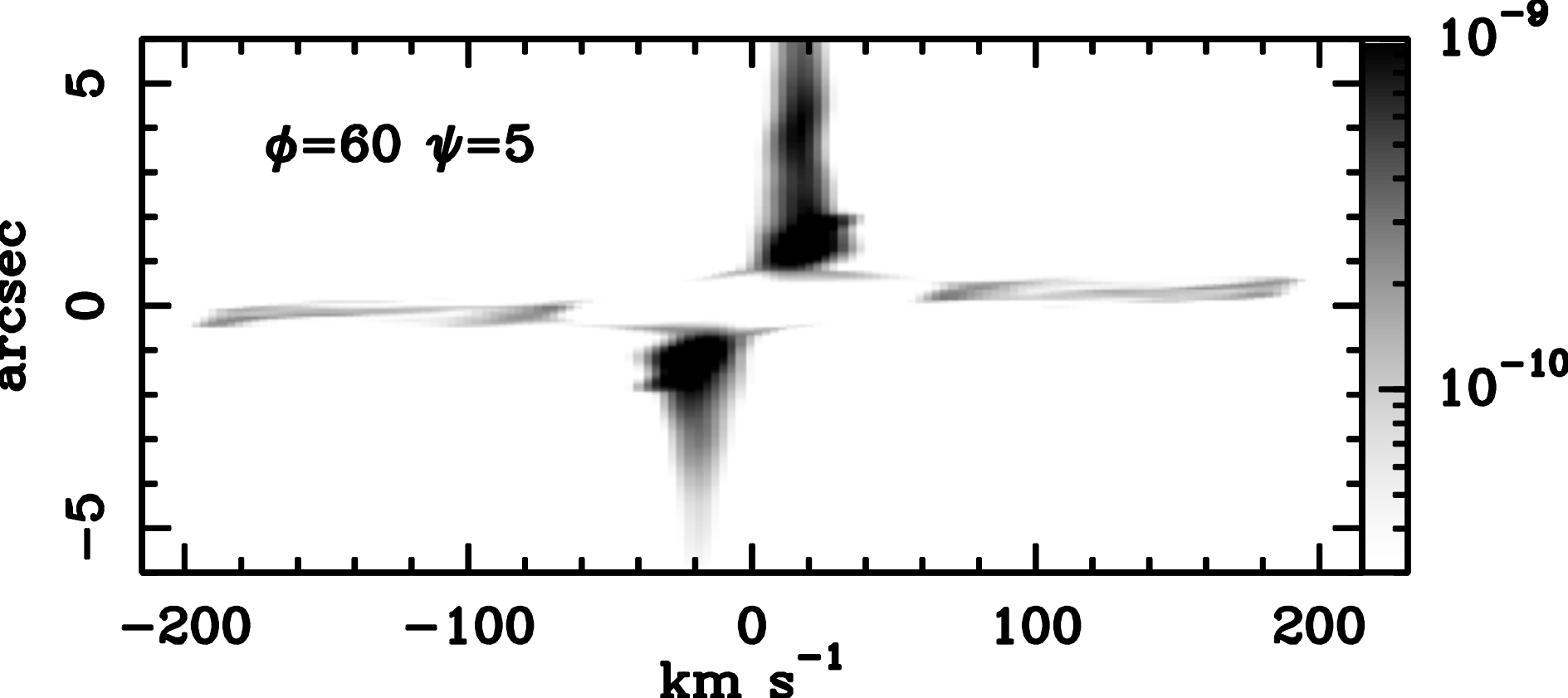}
\includegraphics[width=1.0\linewidth]{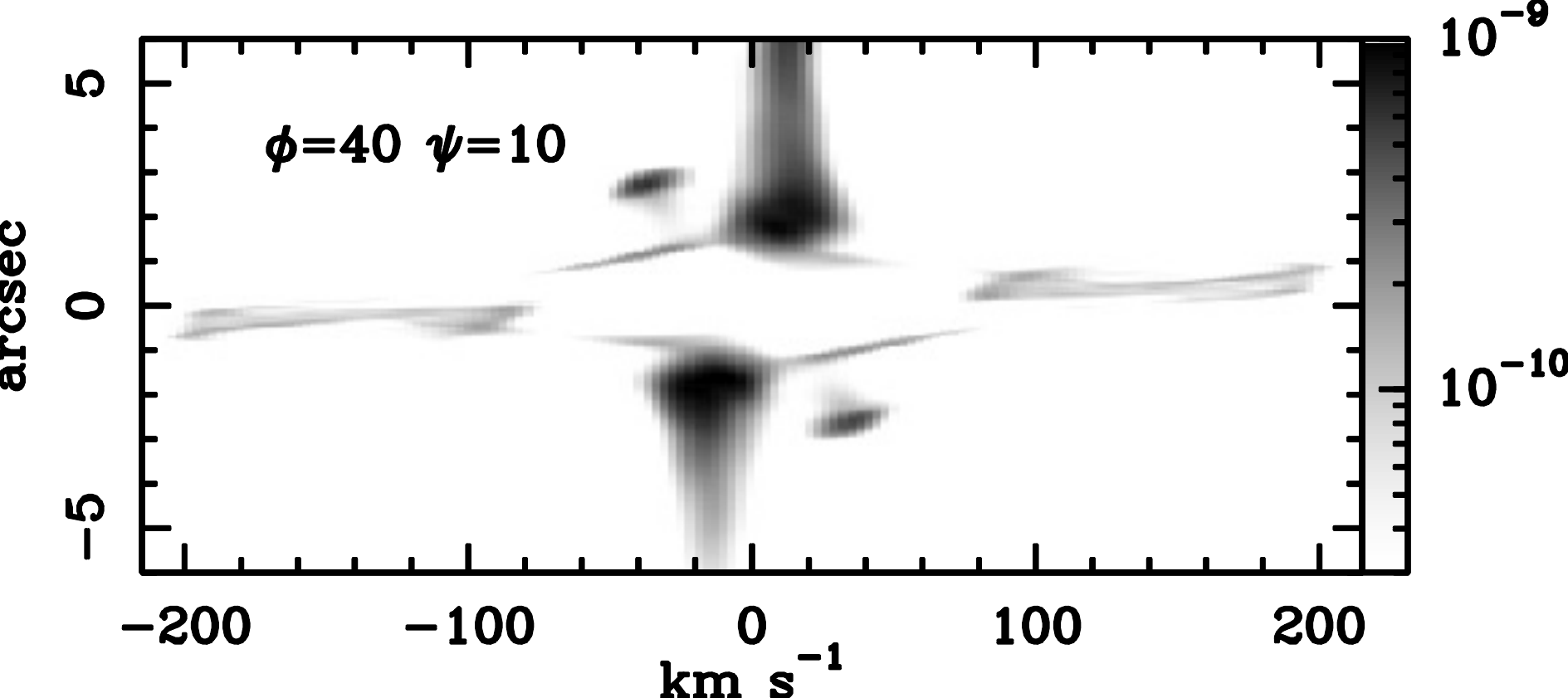}
\includegraphics[width=1.0\linewidth]{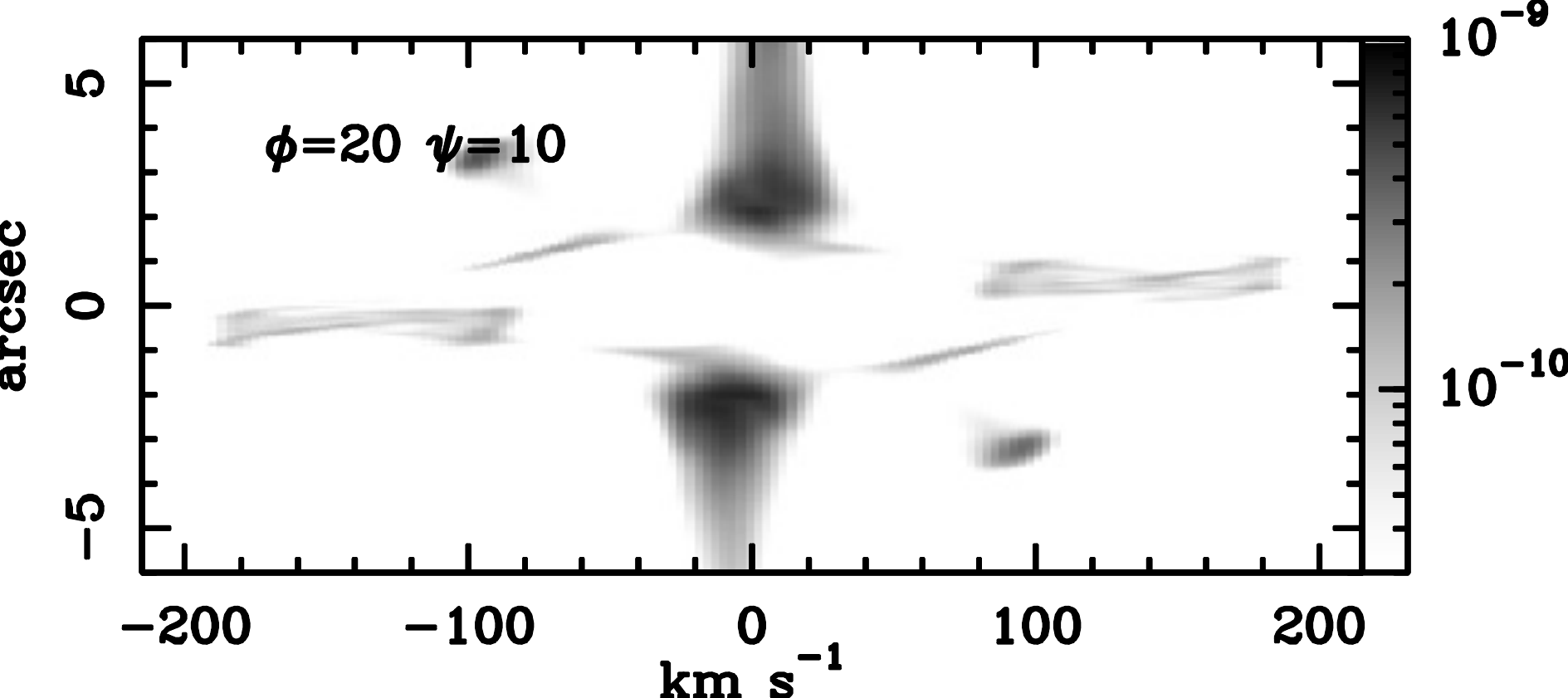}
 \caption{The same as Figure  \ref{fig:pvniixz13phi30Rot5CEN_new4nov_alfa30} but for model 3 and $xz-$projection.}
  \label{fig:pvniixz15phi30Rot5CEN_new4ene_alfa40}
\end{figure}

 \begin{figure}
\centering
   \includegraphics[width=1.05\linewidth]{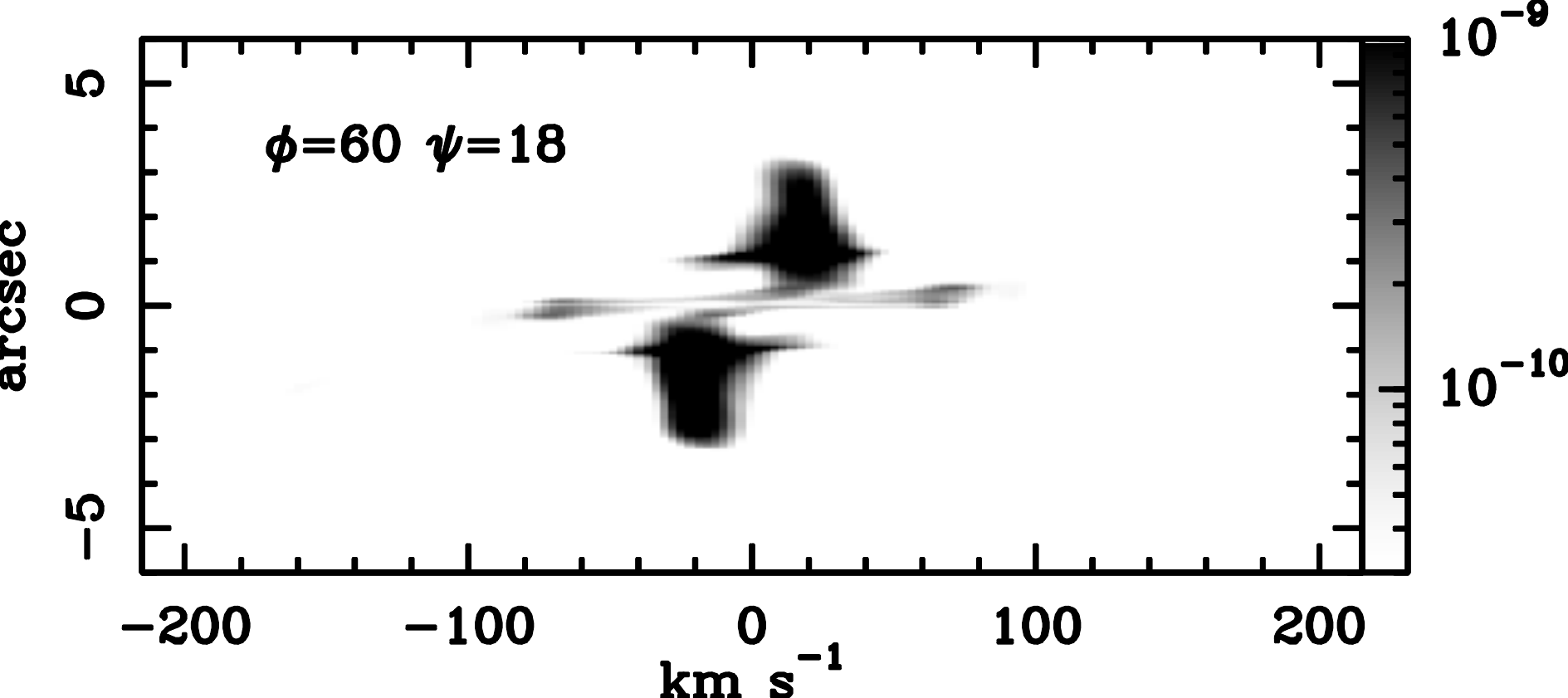}
   \includegraphics[width=1.05\linewidth]{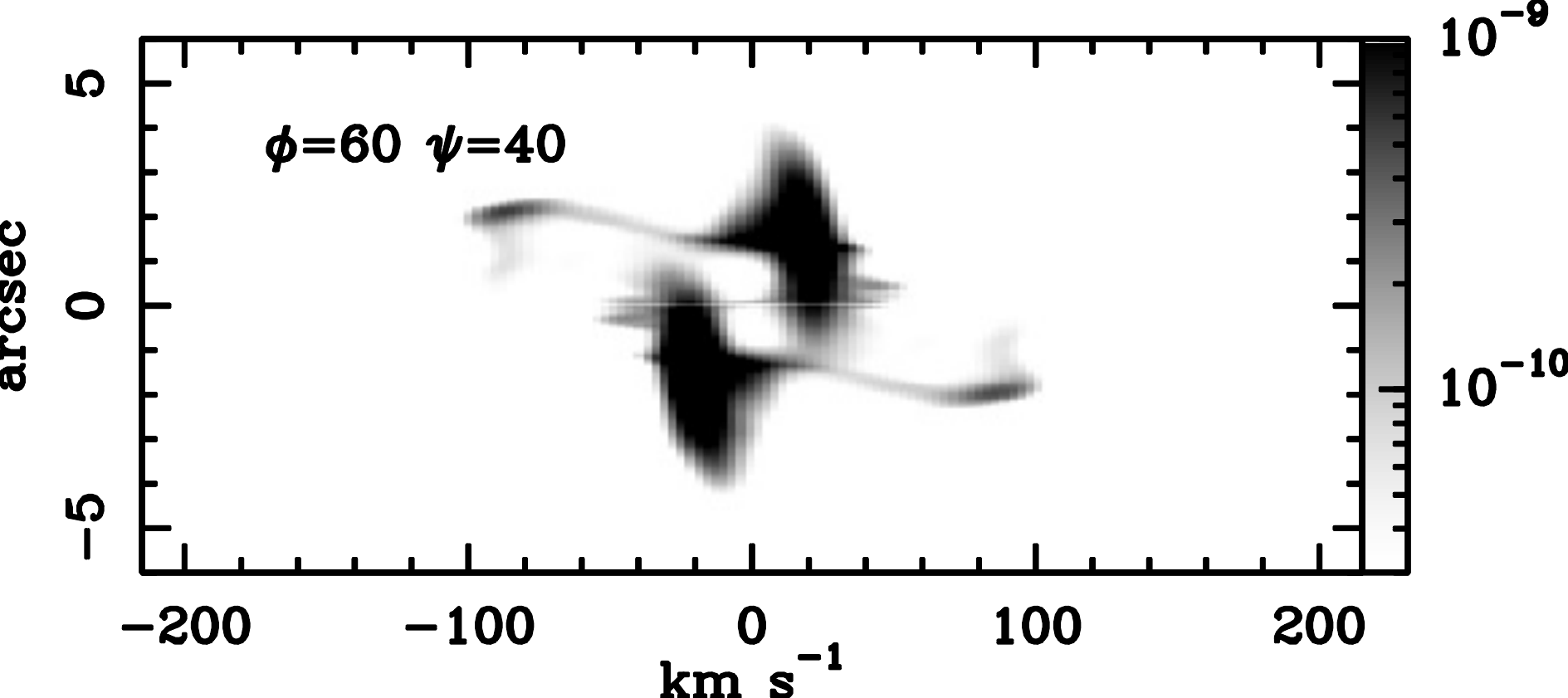}
   \includegraphics[width=1.05\linewidth]{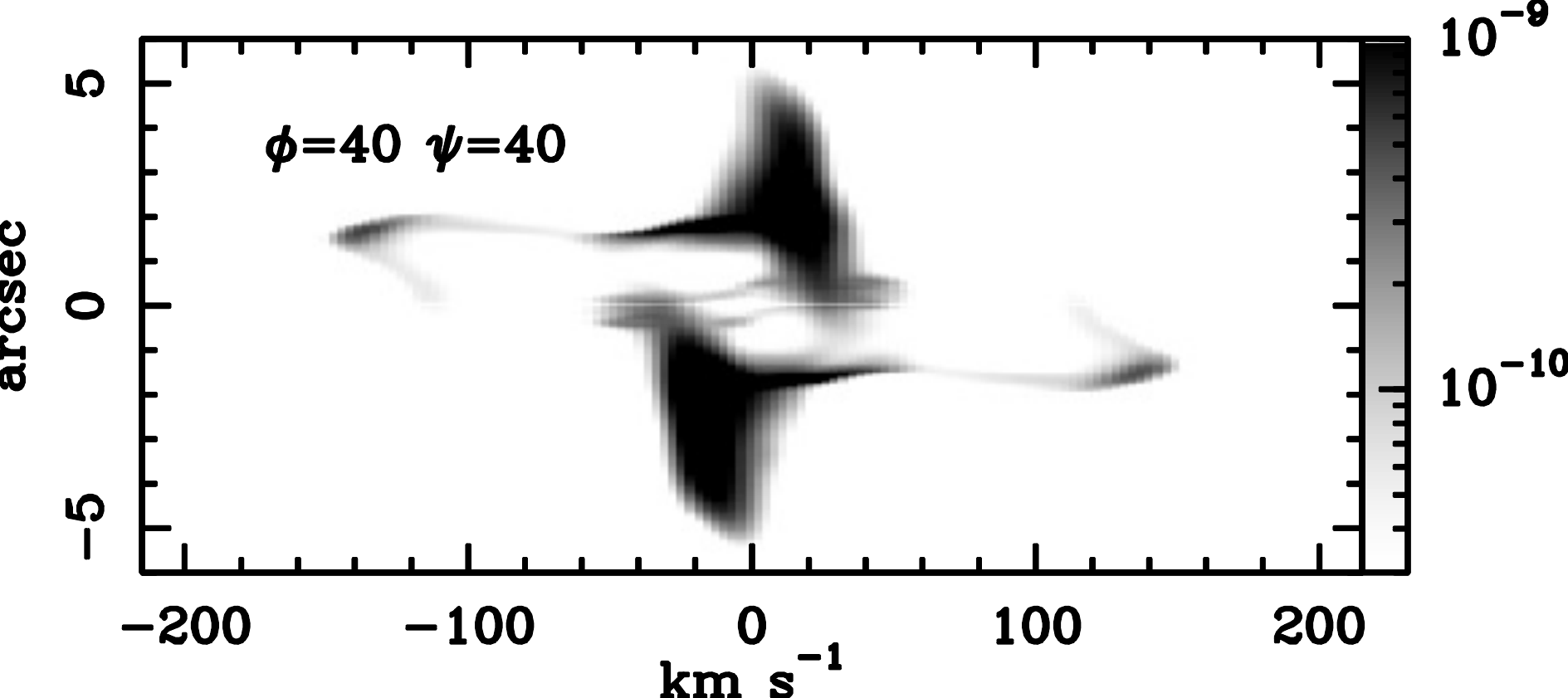}
   
 \caption{The same as Figure \ref{fig:pvniixz13phi30Rot5CEN_new4nov_alfa30} but for model 3 and $yz-$projection. }
  \label{fig:pvniiyz15phi40Rot40Model4}
\end{figure}

\begin{figure}
\centering
   \includegraphics[width=1.0\linewidth]{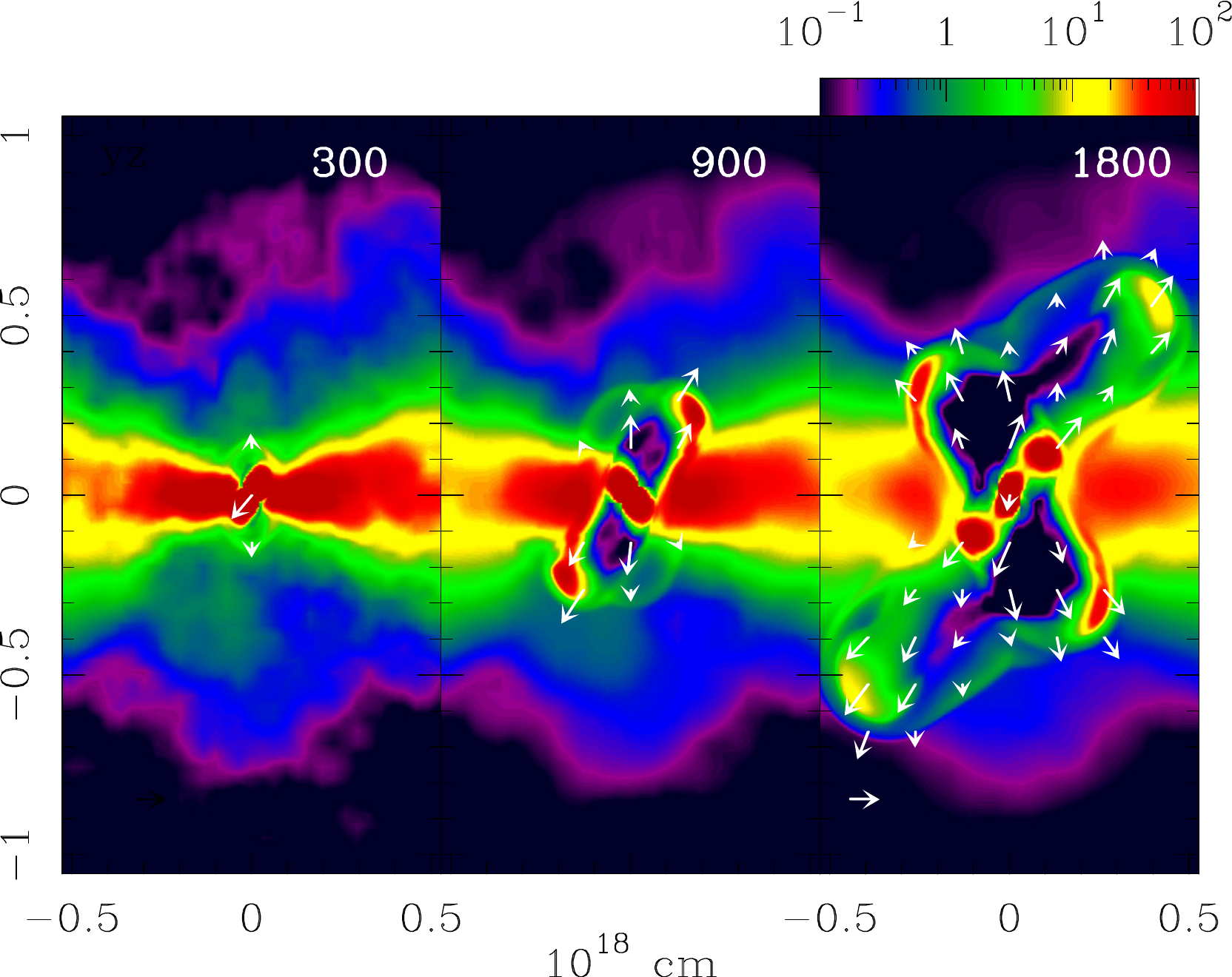}
   
   \includegraphics[width=1.0\linewidth]{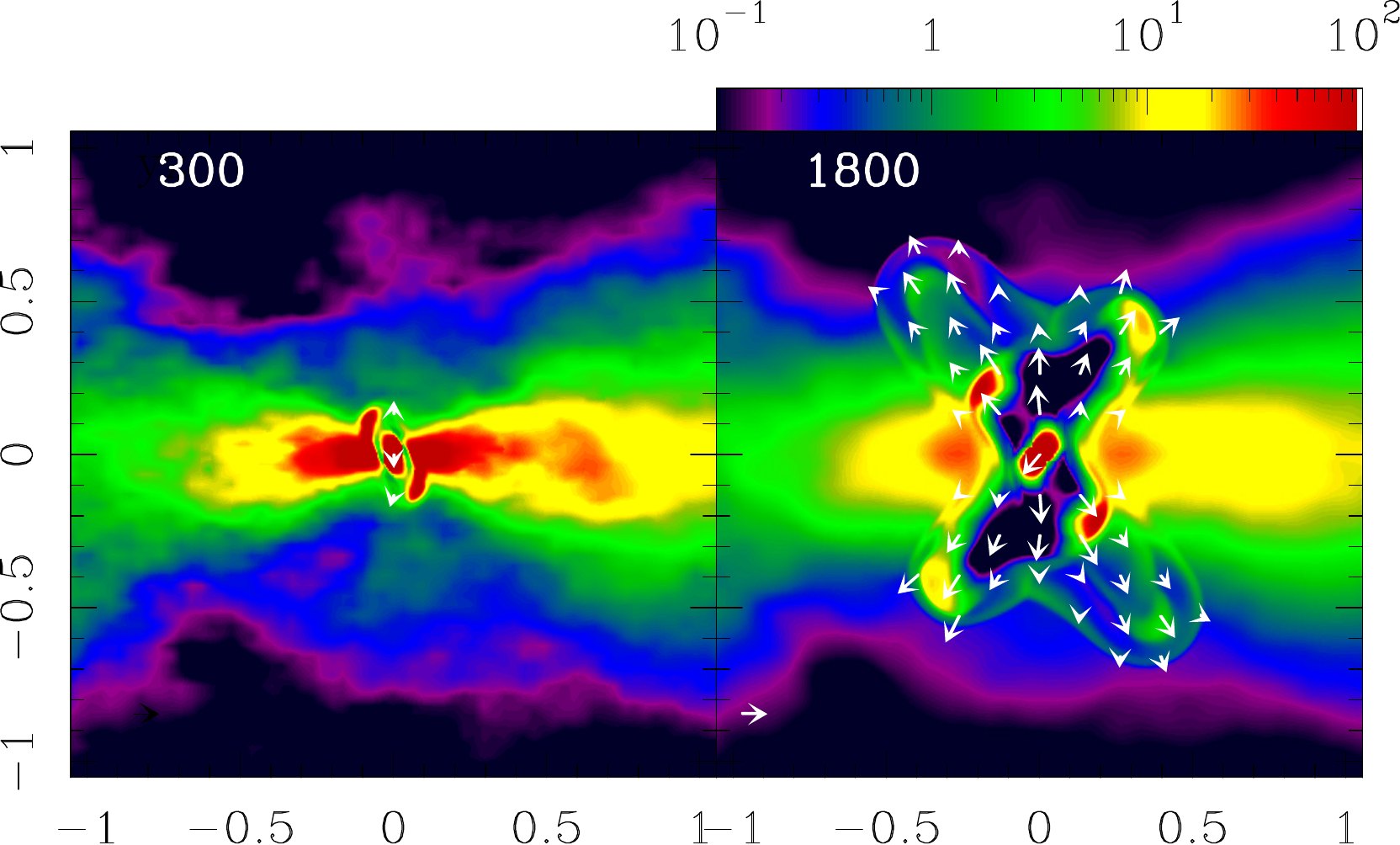}
 \caption{Temporal evolution of the electron density obtained for model 4. Upper panels show the evolution for the $xz$-projection while
 bottom panels display the evolution for the $yz$- projection.}
  \label{fig:deltaV07DenYZ3001800}
\end{figure}

Fig. \ref{fig:pvniixz15phi30Rot5CEN_new4ene_alfa40} shows the synthetic PV diagrams performed for the xz-projection considering angles $\phi$=60$^{\circ}$, 40$^{\circ}$, 20$^{\circ}$ and $\psi$=5$^{\circ}$, 10$^{\circ}$, 10$^{\circ}$. 
Fig. \ref{fig:pvniiyz15phi40Rot40Model4} shows the synthetic PV diagrams performed for the yz-projection considering angles $\phi$=60$^{\circ}$, 40$^{\circ}$ and $\psi$=18$^{\circ}$, 40$^{\circ}$.

For H\,1-67, our observational PV diagram with the slit at 45$^{\circ}$ (Fig. \ref{fig:H1-67pv}, top) is well reproduced by the synthetic PV diagram with $\phi$=60$^{\circ}$ and $\psi$=40$^{\circ}$   (Fig. \ref{fig:pvniiyz15phi40Rot40Model4}, middle).
We reproduced the central condensations corresponding to the dense torus, separated almost 40 km s$^{-1}$ approximately. 
A clear point-symmetric morphology results from the presence of the precession jet. The synthetic PV diagrams show fast outflows up to velocities of $\pm$95 km s$^{-1}$, with a S-shape that is similar to the observations.

We also obtained a good agreement with the PV diagram from observations along P.A. = 0$^{\circ}$  if the
jet axis was tilted at $\phi$=60$^{\circ}$ and $\psi$=18$^{\circ}$ (Fig. \ref{fig:pvniiyz15phi40Rot40Model4} top).  \\

\subsubsection{Model 4, with variable ejections density}

 Instead of time-dependent ejection velocity for the jet, we have computed a model which considers a time-dependent ejection density for the jet, given by Eq. \ref{djet}.
The parameters  for this model are the same as those for Model 3 (the same $\alpha$, $f_p$ and $f_q$). In this case, the jet has constant velocity of 140 km s$^{-1}$. The average density is 150 cm$^{-3}$ and the density variation is 70\%.
The total integration time was of 1800 yr. \\

In Fig. \ref{fig:deltaV07DenYZ3001800}  (top) the xz-projection of the electron density distribution map at 300, 900 and 1800 yr is shown while in the bottom figure, the yz-projection is presented at 300 and 1800 yr. These figures are similar to the ones presented for Model 3, although in this case the knots at high velocity appear denser and more defined.

\begin{figure}
\centering
   \includegraphics[width=1.05\linewidth]{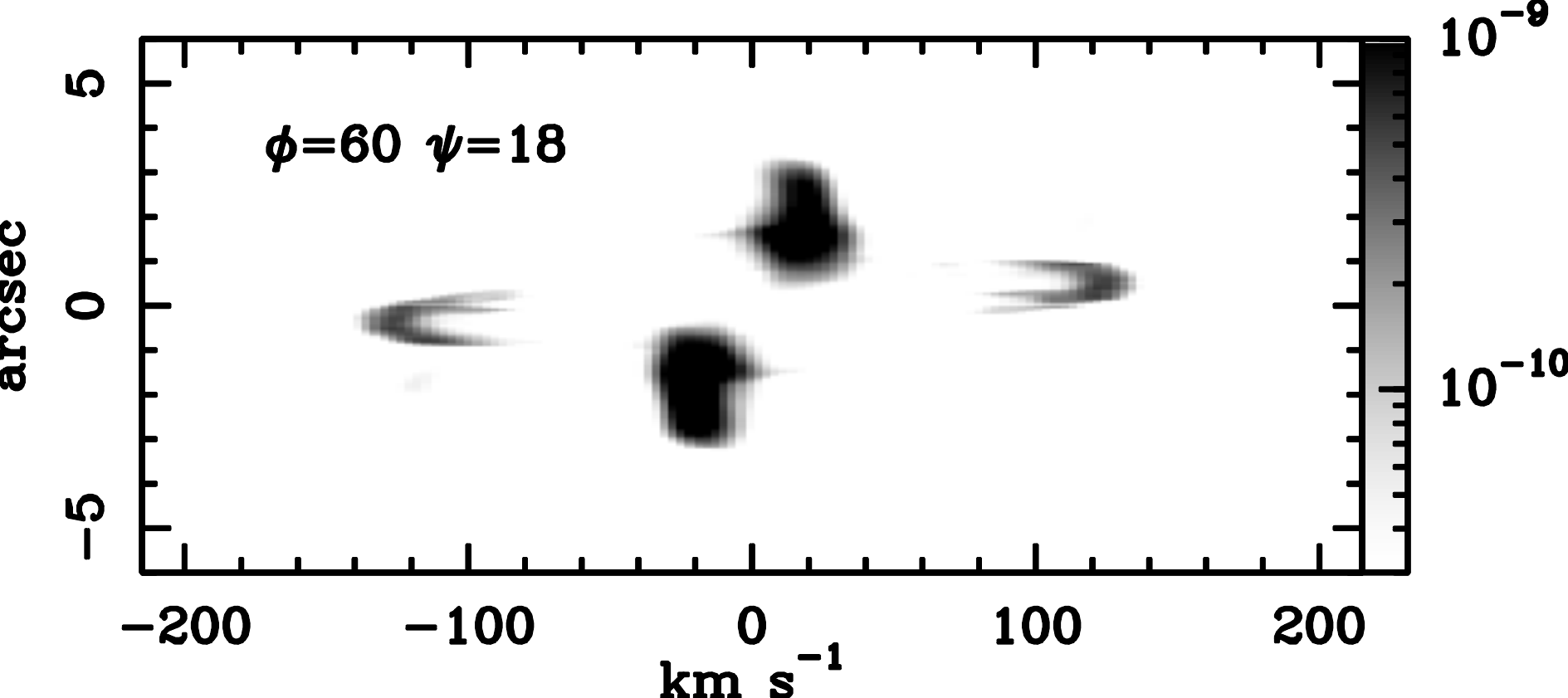}
   \includegraphics[width=1.05\linewidth]{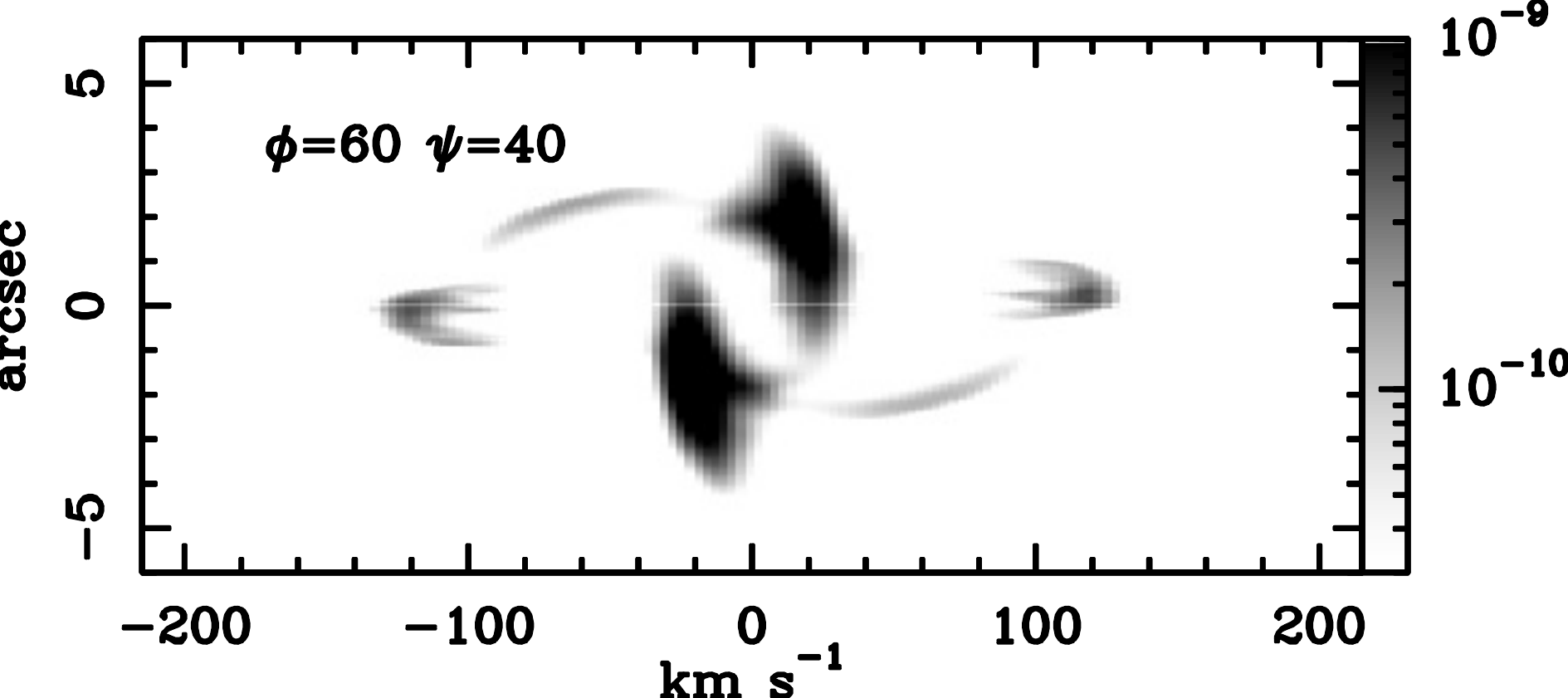}
   
   \includegraphics[width=1.05\linewidth]{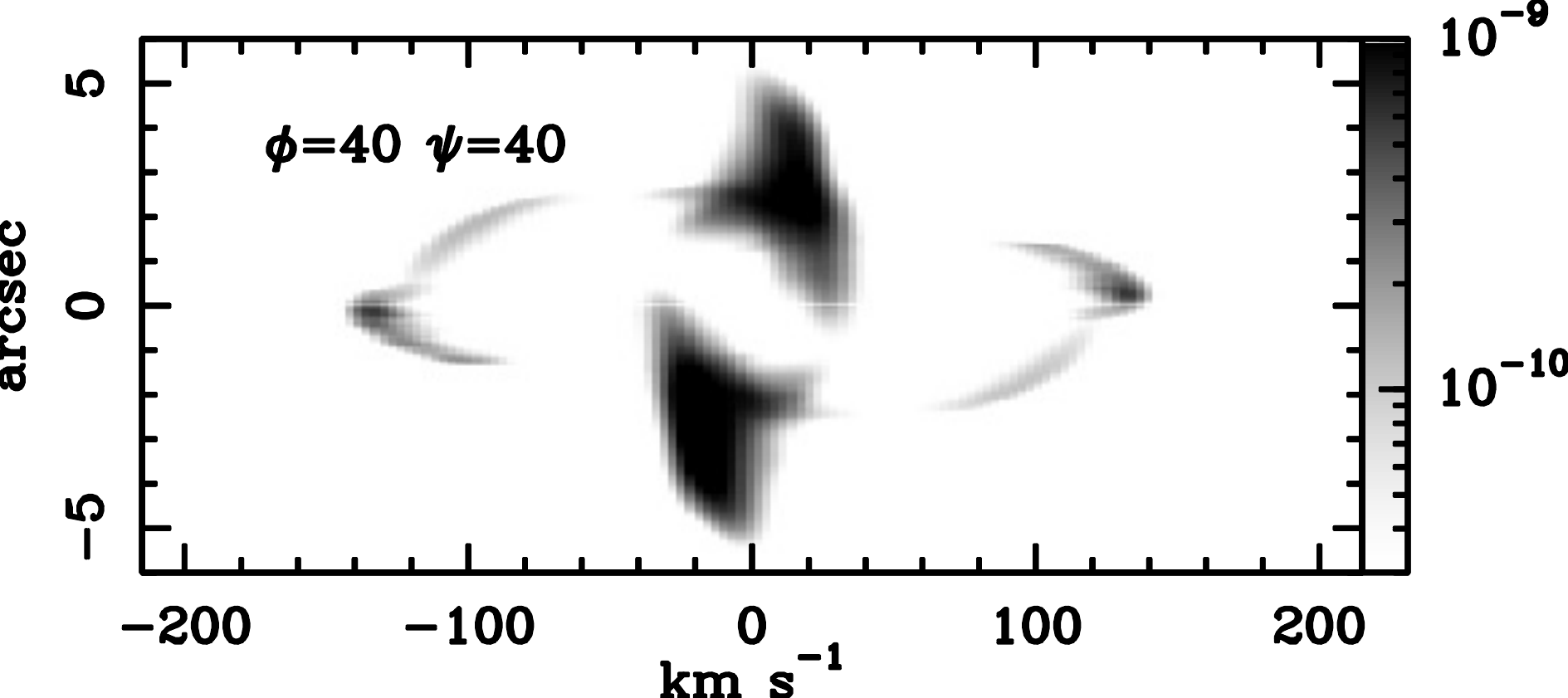}
   
 \caption{The same as Figure \ref{fig:pvniixz13phi30Rot5CEN_new4nov_alfa30} but for model 4 and $yz-$projection. }
  \label{fig:pvniiyz16ModelDensity}
\end{figure}

The synthetic PV diagrams for this model are presented in Fig. \ref{fig:pvniiyz16ModelDensity} where three diagrams are showed with ($\phi$, $\psi$) of (60,18), (60,40) and (40,40). The one  corresponding to $\phi$=60$^{\circ}$ and $\psi$=18$^{\circ}$ reproduces well the observations with the slit at P.A.= 0$^{\circ}$.  The second synthetic PV diagram for $\phi$=60$^{\circ}$ and $\psi$=40$^{\circ}$  reproduces well the observed PV diagram for H\,1-67 for the slit at P.A.= 45$^{\circ}$. In both panels, the knots observed at $\pm$120 km s$^{-1}$ appear well defined in the model. The third synthetic PV diagram show $\phi$=40$^{\circ}$ and $\psi$=40$^{\circ}$, this PV diagram does not reproduce the observations.

\subsubsection{Synthetic Images}

	 We have built  synthetic images corresponding to the H$\alpha$, [N\,{\sc ii}], [S\,{\sc ii}]   emissions from our model 4 for an evolution time of 1800 yr representing the image that could be observed with a very high spatial resolution like that of HST. Fig. 16 shows the synthetic images for the emission of [N\,{\sc ii}]  and [S\,{\sc ii}]. H$\alpha$ is not shown because presents the same structure. It should be noticed that projection does not include the rotation in the sky plane. The brightest emission corresponds to the torus and the filaments are observed above and below, showing a S-shape. These head of filament are moving at $\pm$130 km s$^{-1}$, a velocity highly supersonic therefore they correspond to a shocked gas.

 \begin{figure}[h]
  \centering

\includegraphics[width=0.493\linewidth]{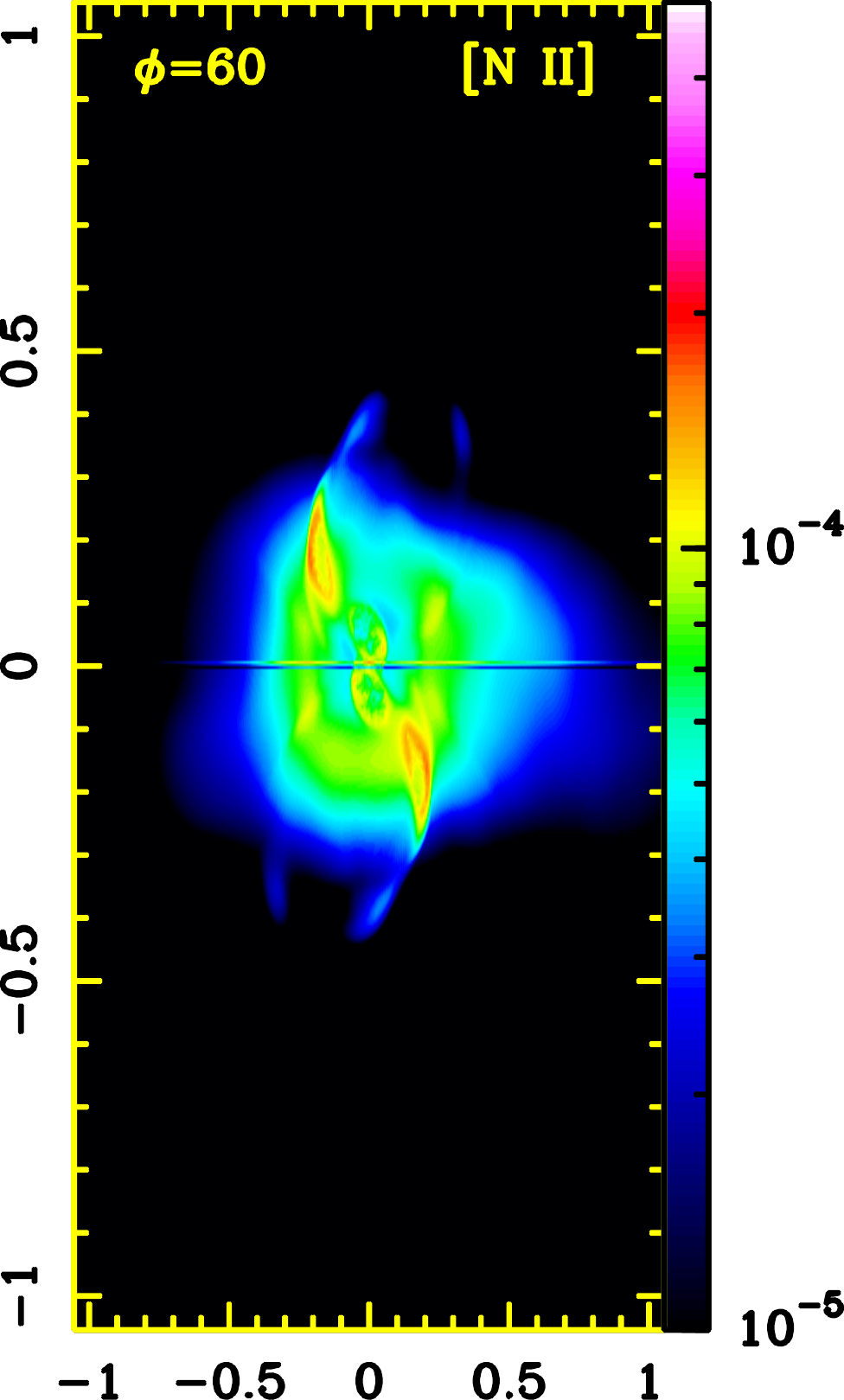}
\includegraphics[width=0.493\linewidth]{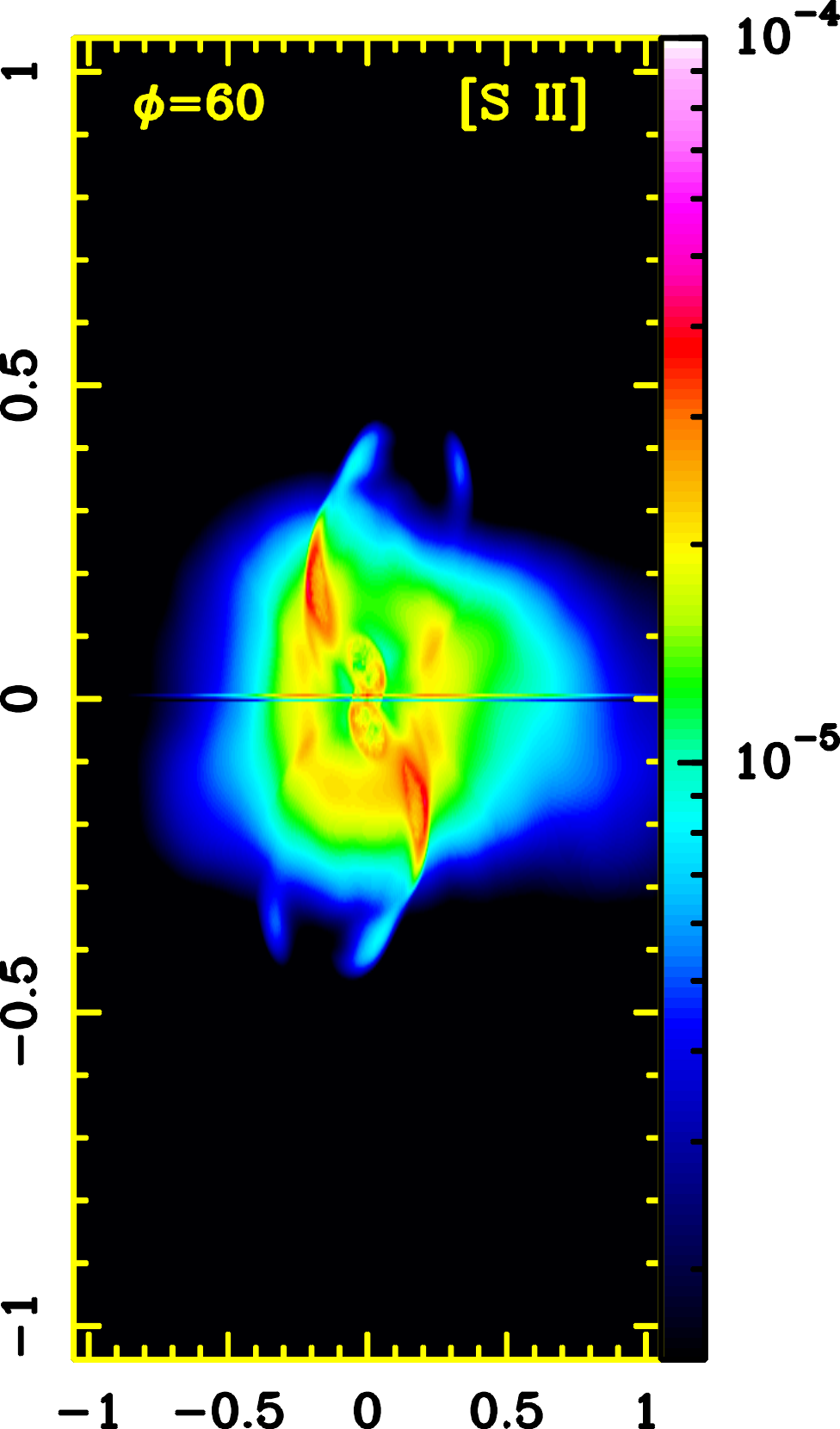}

  \caption{Emission maps with $\phi$ = 60$^{\circ}$ for the $xz$- projection for the model 4 (variable ejections density) at 1800 yr. Right: [N\,{\sc ii}] emission line. Left: [S\,{\sc ii}] emission line. Both axes are given in units of 10$^{18}$ cm and the logarithmic color scale gives the number density in units of cm$^{-3}$.}
  \label{MapasProyXZ}
\end{figure}

\begin{table*}
\centering
 \begin{minipage}{120mm}
  \caption{Input for hydrodynamical simulations}
\begin{tabular}{lllccccc}
\hline \hline
PN G & Name & Stellar & Outflow velocity & A & B  & Spectral & Precession\\
  &  & type &  km s$^{-1}$ &  & & morphology & \\
\hline

006.8+04.1 & M\,3-15  & [WC 4] & 90 & 0.99 & 0.1 & Bipolar & No\\

009.8$-$04.6 & H\,1-67  & [WC 2-3] & 97 & 0.99 & 5 & point-symmetric & Yes\\

006.8+04.1 & M\,1-32  & [WC 4] pec & 180 & 0.9 & 5 & Bipolar & No\\

\hline
\multicolumn{5}{l}{$^a$ model with variable density. In column 6 $\tau_d$ is presented.}
\end{tabular}
\label{tab:hydro-model}

\end{minipage}
\end{table*}

\section{Other objects}
\begin{figure}
  \centering
\includegraphics[width=0.7\linewidth]{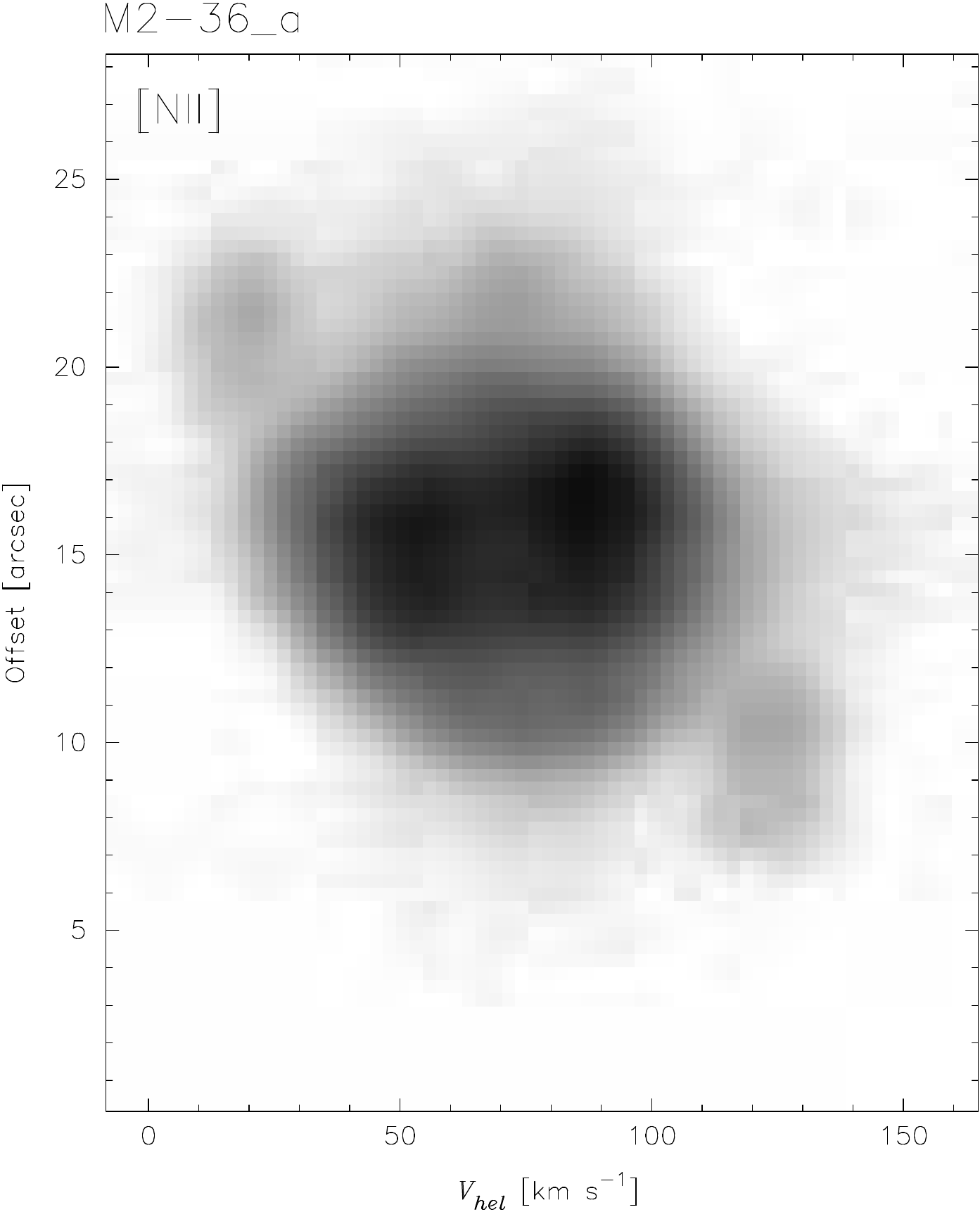}
\includegraphics[width=0.5\linewidth]{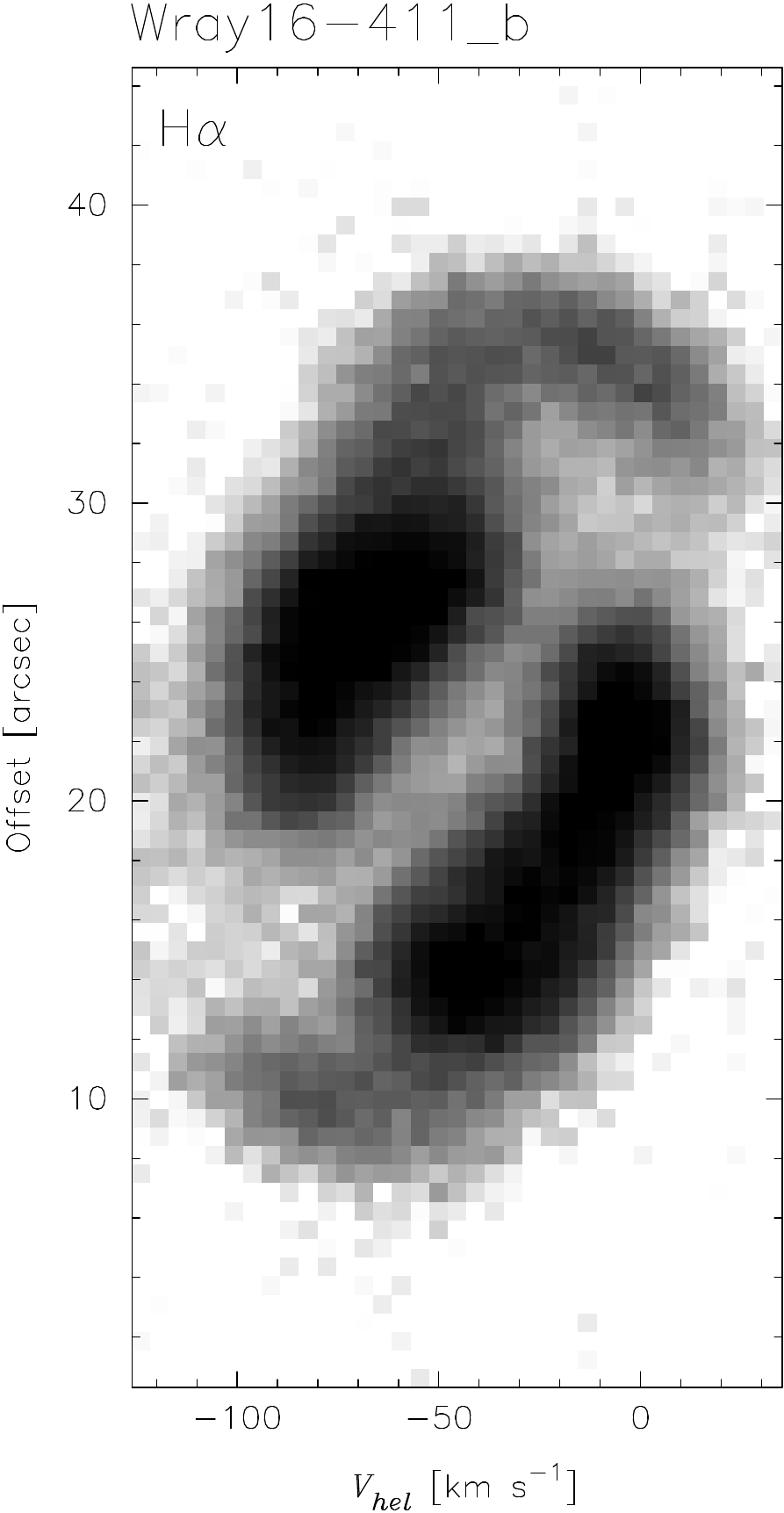}
  \caption{ PV diagrams observed for two PNe, obtained from The SPM
Kinematic Catalogue of Planetary Nebulae. Top: PN M 2-36. We observe two bright knots at the centre and two faint knots with S-shape reaching high velocities. Bottom: PN Wray 16-411. We observed two bright elongated condensations at the centre and high-velocity outflows with S-shape. These PNe are good candidates
for our hydrodynamical model. }
  \label{fig:PSPNpv}
\end{figure}

Using the available spectra in The SPM Kinematic Catalogue of Galactic Planetary Nebulae \citep{Lopez2012}, we have found
two PNe, M\,2-36  (PN G 003.2$-$06.2) and Wray 16-411 (PN G 353.7$-$12.8), with similar spectral characteristics to H\,1-67. In their spectra they present a S-shape morphology with knots or filaments at high velocity (see Fig.\ref{fig:PSPNpv}). 
Some of their characteristics are listed in Table 2. 
M\,2-36 is located towards the  galactic bulge, showing a radial velocity of 57 km s$^{-1}$ \citep{RicherSuarez2017} and has an ADF\footnote{ADF is the Abundance Discrepancy Factor given by the ratio of ionic abundances derived from optical recombination lines and collisionaly excitation lines} of 6.9 \citep{LiuLuo2001}. Due to this large ADF, this PN is a candidate to have a binary central star \citep{CorradiGarcia2015}. Its heliocentric distance calculated by \citet{Zhang1995} is 6.99
 kpc. The total nebular abundance ratios for this PN are N/O = 0.95 and He/H = 0.135 \citep{LiuLuo2001}, therefore, this object is a helium and nitrogen rich nebula, classified  as a Peimbert Type I PN.

Wray 16-411  is located at a distance of 6.14 kpc \citep{FrewParker2016}. Its heliocentric radial velocity is $-$50 km s$^{-1}$ \citep{BeaulieuDopita1999}. \citet{WeidmannSchmidt2016} mention that this object could be the galaxy IRAS 18232−4031 due to its spiral morphology, but the spectral data in the SPM Catalogue \citep{Lopez2012}  show clearly that it is a planetary nebula.

Our models could help us to reproduce these PNe with the same elements used here, a precessing jet with time-dependent velocity or density ejection,  and a broken torus, but using other values for $f_p$, $f_q$ and $\alpha$. This work is in process. 

\section{Discussion and Conclusions}
 In the literature there are several point-symmetric PN that present regular knotty ejections such as Fg 1 and Hen 2-90. 
\citet{SahaiNyman2000} analyzed the case of Hen 2-90 which shows knotty filaments with many ejections along a straight line at both sides of a toroidal structure, in this case the jet seems to be not precessing. \citet{SahaiNyman2000} estimated an ejection time interval of the knots between 100-120 yrs for this nebula. Another case already mentioned  is Fg\,1, where the S-shape is very extended (the filaments span 2 arcmin to either side of the PN) and knotty. Its kinematical structure was  analyzed by \citet{LopezMeaburn1993}. It is found that the knots were not emitted as regularly as in Hen 2-90. In this case the jet is precessing and from the figures presented by \citet{LopezMeaburn1993} we estimated that the ejection time interval between knots is approximately larger than 1000 yrs.\\   

From the above it is apparent that the existence of knotty jets is relatively common in point-symmetric PNe. In many cases the knots are emitted in time intervals of several hundreds of years and present masses between 10$^{-6}$ and 10$^{-3}$ M$_\odot$.\\

The knotty filaments presented by point-symmetric PNe  are related to the mass-loss processes 
in the central stars. It is clear that some AGB and post-AGB stars eject material not continuously but in regular (or irregular) mass-loss events. In the case of point-symmetric PNe, the ejection corresponds to collimated  jets moving in different (opposite) directions, sometimes precessing and giving origin to S-shape structures.\\

In addition to these point-symmetric objects, there is a fraction of PNe and proto-PNe showing external  structures in the form of rings and arcs around their central stars, occasionally regularly spaced. By analyzing a large number of images searching for external rings and arcs, \citet{RamoslariosSantamaria2016} reported that about 8\% of PNe present this type of structures. These authors investigated the spacing between rings and arcs and their number in 29 objects. A few of the PNe presenting rings and arcs, also present filaments and point-symmetric components like NGC\,6543, a PN with a very complex morphology \citep[][]{BalickWilson2001, MirandaSolf1992}.

\citet{RamoslariosSantamaria2016} estimated that the averaged time-lapse between
rings and arcs, in the PNe of their sample, is in the range from 500 to 1200 yr, which is similar to the time interval between ejections in the knotty point-symmetric PNe. 
It is possible that the mechanisms of ejection of both type of structures are similar. 
However the expansion velocities of the collimated outflows are in general large (several tenths of km s$^{-1}$ and in some cases larger than 1000 km s$^{-1}$) while the expansion velocities of rings and arcs (basically unknown) are assumed to be the typical expansion velocity for an AGB envelope, of about 10 km s$^{-1}$. Knotty jets and rings are repetitive phenomena occurring in PNe, and they present similar time scales  but they correspond to different epochs of ejections, rings seem to be produced at the end of the AGB phase, while periodic jets seem to be produced when the PN is already formed. The morphology of both types of structures is also very different, ring and arcs being spherically symmetric, while collimated jets are filamentary structures.
Therefore, except for the time scales, both phenomena seem unrelated.

 Comparing the point-symmetric PNe mentioned above, we found that many of them have a disk or torus in the centre and a precessing jet flowing through the poles. Both components
are necessary to produce this kind of morphology. The structure of the knotty collimated outflows could be due to an ejection with time-dependent velocity or time-dependent density.\\

In this work we have carried out several 3D hydrodynamical models with the code Yguaz\'u, in order to explore and understand the processes that give origin to S-shaped planetary nebulae. It is assumed that such a dynamical structure  is  a consequence of the effect of a highly collimated precessing jet, as it was mentioned above. Therefore our hydrodynamical simulations consist of high-velocity collimated outflows in a medium
given by the noisy wind of an AGB, simulating the density distribution of a dense torus. This jet changes its direction forming a cone of precession and it  could be treated as a time-dependent  velocity ejection or a time-dependent  density ejection. 

Our main results from these models can be summarised as: A precession jet is required to obtain such a S-shape morphology.  The semi-aperture angle of the precession cone should be equal or larger than 30$^{\circ}$ for obtaining this structure and the point-symmetric morphology is very evident when the angle is 40$^{\circ}$. In addition, the factor of precession ($f_p$) and the number of ejections ($f_q$) in each turn will determine the structure of the S-shaped filaments.\\

Two models are presented here, the first one with several turns ($f_p = 4$) and ejections ($f_q = 2$), and a second one with smaller number of turns ($f_p = 1.5$) and ejections ($f_q = 2$). The torus included in the models helps to produce  a bipolar nebula, but  has no effect on the S-shape. Only the precessing jet produces such a shape. The projection of this structure on the sky plane can be compared to real observations.\\

The PV diagrams computed for the models, in comparison with observed PV diagrams of the  PN H\,1-67, allowed us to produce a more refined model for this compact nebula.
 In this work we show that H\,1-67  presents a spectroscopic point-symmetric morphology, even when its image does not show this morphology.  
For this nebula we obtained high resolution spectra, locating the slit along the symmetry axis of the nebula (P.A. = 45$^\circ$), finding that they present  two bright condensations in the centre corresponding to a broken torus and a S-shape filament with knots at high velocity. Also, when the slit is located with a P.A = 0$^\circ$, this S-like morphology is not evident. From the models described above, we calculated new models varying different parameters trying to reproduce such a structure. \

Synthetic  PV diagrams were generated, considering different inclinations of the models with respect to the line of sight and on the plane of the sky.  From comparing these synthetic PV with the observed PV diagrams of H\,1-67, we found that our Model 3 with time-dependent velocity ejections, semi aperture angle of 40$^{\circ}$, $f_p = 1.5$, $f_q = 4$, and Model 4 with a time-dependent density ejections and the same parameters,   are the ones that best reproduce the characteristics of the PV diagrams for this object, after an integration time of 1500 and  1800 yr respectively. According to our models, each ejection is launched every 345 yr (model 3) and 305 yr (model 4).\\

 Both models, 3 and  4, show two bright condensations with velocities close to 0 km s$^{-1}$, which are associated with the toroidal material. In the orientation $\phi$ = 60$^{\circ}$, $\psi$ = 40$^{\circ}$ the models  show the observed S-shape ending in two knots at $\pm$100 and $\pm$120 km s$^{-1}$ respectively, these PV diagrams reproduce our observations of H\,1-67 with the slit at 45$^{\circ}$. In the orientation $\phi$ = 60$^{\circ}$, $\psi$ = 18$^{\circ}$,  two knots are seen at velocity of the order of $\pm$100 and $\pm$120 km s$^{-1}$ respectively, and no S-shape is observed. These PV diagrams reproduce our observations of H\,1-67 with the slit at 0$^{\circ}$.\\

 By using different sets of parameters, these hydrodynamical models could be applied to reproduce other compact PNe that present PV diagrams showing a S-shape structure, such as Wray\,16-411 and M\,2-36.\\

In summary our hydrodynamical model is a simple one.  It consists of a bipolar jet  moving inside an anisotropic AGB wind (with high density at the equator). Such a model helped us to reproduce the PV diagrams of the spectroscopic bipolar PNe M\,1-32 and M\,3-15 \citep{RechyGarcia2017}. In this work we show that adding precession and time-dependent ejection velocity or density we can reproduce the point-symmetric PV diagram of H\,1-67.\\ 

\vskip 0.2mm

 The PN H1-67 analyzed in this work presents a point-symmetric morphology seen from spectroscopic observations  but not in an optical image due to the collimated outflow is projected onto the torus which is much brighter. Possibly the point-symmetric structure of H\,1-67 could be observed in a very high-resolution image with $HST$ such as we show in our synthetic image (Fig. 16). In this paper, the idea that point-symmetric PNe results from the periodic ejection of dense material by a precessing jet is corroborated.\\ 

Some PNe such as Hen 2-11 \citep{JonesBoffin2014}, Hen 2-155 \citep{JonesBoffin2015}, NGC 6326 and NGC 6778 \citep{MiszalskiJones2011} appear to possess 
filaments and jet-like structures around a binary stellar system. Therefore the point-symmetric structure that presents H\,1-67 could be a consequence of a binary system. This should be investigated in the future.

\section*{Acknowledgments}
 We thank the daytime 
and night support staff at the OAN-SPM, Gustavo Melgoza, Salvador Monroy, and Felipe Montalvo, for facilitating and helping to obtain our observations. Helpful comments by Dr. Jos\'e Alberto L\'opez, Dr. Michael G. Richer, Dr. Christophe Morisset and Dr. Gloria Delgado-Inglada 
are deeply acknowledged. We thank the anonymous referee for her/his very useful suggestions and comments, which help us to improve the previous version of this manuscript.
We thank Enrique Palacios (C\'omputo-ICN) for mantaining
the Linux servers where the hydrodynamical simulations were carried out.  This work
received finantial support from DGAPA-PAPIIT grants IN103117, IG100218
and  CONACyT grant 241732. J.S.R.-G. acknowledges scholarship from
CONACyT-M\'exico.

\bibliography{ref}{}
\bibliographystyle{mn2e}

 \appendix
\section{Distance of H1-67 from GAIA}
During the revision of this manuscript The  GAIA Data Release 2 (GAIA DR2) have been published and there  a parallax of 1.71 mas $\pm$ 0.50 mas is reported for H1-67.
This would indicate a heliocentric distance between 450 and 830 pc (585 pc $+240 \atop -130$).
This distance is 10 times smaller than the 5880 pc reported by \citet{FrewParker2016}  which represents a parallax of 0.17 mas.
On the other hand (\citet{Stanghellini2010}  report 
a distance of  7860 pc for this object, equivalent to a parallax of  0.13 mas).
The observed angular diameter of H\,1-67 is 5.6 arcsec, equivalent to 0.15 pc at a distance of 5880 pc.
With the distance of 585 pc from GAIA DR2, the physical diameter of H\,1-67 would  0.015 pc,
and the age, estimated as $t = radius/V_{exp}$ would be only 150 to 200 yr instead of 1500 to 2200 yr derived from the distance by \citet{FrewParker2016}. With these parameters H1-67 would be  the youngest PN known in the Galaxy. 
\\

Other parameter that is largely affected with this change of distance is the ionised mass. 
Calculating the ionised mass  as M=$4/3 \pi R^3 n m_{H}$   it is obtained  0.00619 M$_{\odot}$ with GAIA DR2 distance, and if the distance is 5.88 kpc (Frew) the ionized mass is 0.15 M$_{\odot}$.
\\

The mass can be also calculated from the H$_{\beta}$ flux  with the expression

$$M_i(M_{\odot}) = 8.07 x 10^{11} I(H_{\beta}) d^2 t_e^{0.88}(1+4y)/n_e$$
\\
it is obtained 0.001 M$_{\odot}$ for a distance of 585 pc, while the mass for a distance of 5.88 kpc
is of 0.1 M$_{\odot}$, which is much more reasonable. We have assumed  I(H$_{\beta}$) =  1.5$\times$10$^{-12}$  erg cm$^{-2}$ s$^{-1}$ from \citet{EscuderoCosta2004}, multiplied by 3 to take into account that the slit was 2 arcsec  wide 
and the object diameter is about 6 arcsec.
\\

All seems to indicate that the parallax reported by GAIA DR2 is incorrect for this object, because a too small radius, 
age, and ionised mass is derived with such a small distance.
The parallax from GAIA DR2 could be incorrect for H1-67, possibly because it is an extended object with knots and filaments.
The central star is too faint and possibly it was no detected by GAIA.

\bsp

\label{lastpage}

\appendix

\end{document}